\documentclass[aps,prc,twocolumn,floatfix,superscriptaddress,showpacs,nofootinbib]{revtex4-1}

\usepackage{graphicx,amsmath,amssymb,bm,color}

\DeclareGraphicsExtensions{ps,eps,ps.gz,frh.eps,ml.eps}
\usepackage[english]{babel}
\usepackage{fmtcount}
\usepackage{graphicx}
\usepackage{amsmath, amsfonts, amssymb, amsthm, amscd, amsbsy, amstext}
\usepackage{marvosym}
\usepackage{braket,xspace}
\usepackage{mathrsfs}
\usepackage{booktabs}
\usepackage{upgreek}
\usepackage{array}
\usepackage[makeroom]{cancel}
\usepackage{wasysym}

\usepackage{blindtext}
\usepackage{url}
\usepackage{hyperref}
\hypersetup{
    pdfsubject=Paper,
    pdfkeywords={nuclear physics} {dilute Fermi gas} {effective field theory},
    unicode = true,
    breaklinks = true,
    colorlinks = true,
    linkcolor = blue,
    citecolor = blue,
    menucolor = blue,
    citecolor = blue,
    urlcolor = blue
}

\DeclareMathOperator{\e}{\operatorname{e}}

\makeatletter
\newcommand{\overleftrightsmallarrow}{\mathpalette{\overarrowsmall@\leftrightarrowfill@}}
\newcommand{\overrightsmallarrow}{\mathpalette{\overarrowsmall@\rightarrowfill@}}
\newcommand{\overleftsmallarrow}{\mathpalette{\overarrowsmall@\leftarrowfill@}}
\newcommand{\overarrowsmall@}[3]{
  \vbox{
    \ialign{%
      ##\crcr
      #1{\smaller@style{#2}}\crcr
      \noalign{\nointerlineskip}%
      $\m@th\hfil#2#3\hfil$\crcr
    }%
  }%
}
\def\smaller@style#1{%
  \ifx#1\displaystyle\scriptstyle\else
    \ifx#1\textstyle\scriptstyle\else
      \scriptscriptstyle
    \fi
  \fi
}
\makeatother
\newcommand{\te}[1]{\overleftrightsmallarrow{#1}}
\newcommand{\tel}[1]{\overleftsmallarrow{#1}}
\newcommand{\ter}[1]{\overrightsmallarrow{#1}}

\def\XXint#1#2#3{{\setbox0=\hbox{$#1{#2#3}{\int}$ }
\vcenter{\hbox{$#2#3$ }}\kern-.6\wd0}}

\newcommand{\kF}{k_{\rm F}}
\newcommand{\X}{\mathcal{E}}
\newcommand{\x}{\varepsilon}

\newcommand{\breq}{\nonumber \\}
\newcommand{\bs}{\mathbf}
\newcommand{\F}{\text{F}}
\newcommand{\D}{\mathcal{D}}
\newcommand{\eps}{\varepsilon}
\newcommand{\veps}{\epsilon}

\DeclareMathOperator{\pr}{pr}
\newcommand{\given}{\,|\,}
\newcommand{\iid}{\text{i.i.d.}}
\newcommand{\cbar}{\bar{c}}

\begin {document}

\title{Effective field theory for dilute Fermi systems at fourth order}

\author{C. Wellenhofer}
\email[E-mail:~]{wellenhofer@theorie.ikp.physik.tu-darmstadt.de}
\affiliation{Technische Universit\"{a}t Darmstadt, Department of Physics, 64289 Darmstadt, Germany}
\affiliation{ExtreMe Matter Institute EMMI, GSI Helmholtzzentrum f\"{u}r Schwerionenforschung GmbH, 64291 Darmstadt, Germany}

\author{C. Drischler}
\email[E-mail:~]{drischler@frib.msu.edu}
\affiliation{Facility for Rare Isotope Beams, Michigan State University, MI 48824, United States of America}
\affiliation{Department of Physics, University of California, Berkeley, CA 94720, United States of America}
\affiliation{Nuclear Science Division, Lawrence Berkeley National Laboratory, Berkeley, CA 94720, United States of America}

\author{A. Schwenk}
\email[E-mail:~]{schwenk@physik.tu-darmstadt.de}
\affiliation{Technische Universit\"{a}t Darmstadt, Department of Physics, 64289 Darmstadt, Germany}
\affiliation{ExtreMe Matter Institute EMMI, GSI Helmholtzzentrum f\"{u}r Schwerionenforschung GmbH, 64291 Darmstadt, Germany}
\affiliation{Max-Planck-Institut f\"{u}r Kernphysik, Saupfercheckweg 1, 69117 Heidelberg, Germany}

\begin{abstract}
We discuss high-order calculations in perturbative effective field theory 
for fermions at low energy scales.
The Fermi-momentum or $k_{\rm F} a_s$ expansion for the ground-state energy 
of the dilute Fermi gas is calculated to fourth order, both in cutoff 
regularization and in dimensional regularization. 
For the case of spin one-half fermions we find from a Bayesian analysis that 
the expansion is well-converged at this order for ${| k_{\rm F} a_s | \lesssim 0.5}$.
Furthermore, we show that Pad{\'e}-Borel resummations can improve the convergence 
for ${| k_{\rm F} a_s | \lesssim 1}$.
Our results provide important constraints for nonperturbative calculations of 
ultracold atoms and dilute neutron matter.
\end{abstract}

\maketitle

\section{Introduction}

Over the last two decades, striking progress in quantum many-body physics has 
been achieved especially through well-controlled experiments with ultracold atoms 
and the development of efficient computational methods.
Parallel to this, the conception 
of effective field theory (EFT) 
has equipped advanced many-body calculations with a firm theoretical basis.
Here, we make a new contribution to these advances 
by providing analytic EFT results at high orders for a central problem of many-body theory and experiment:
the ground-state energy of the dilute Fermi gas.

Effective field theory is deeply connected with the notion of universality~\cite{Braaten:2004rn}, 
for which the dilute Fermi gas is a classic example. 
This universal many-body system describes both the physics 
of cold atomic gases as well as that of the dilute nuclear matter present in the crust of neutron stars.
In ultracold-atom experiments, Feshbach resonances allow one to tune the interaction 
strength 
via the application of external fields.
This makes it possible to probe low-density Fermi systems over a wide range of many-body dynamics,  
in particular at the unitary limit of infinite scattering length 
and through the BCS-BEC crossover~\cite{Ku563,Navon729,RevModPhys.80.885,RevModPhys.80.1215,RevModPhys.82.1225}.
Moreover, continuous progress with quantum Monte Carlo (QMC) methods~\cite{RevModPhys.80.1215,PhysRevA.84.061602,Gandolfi:2015jma} 
has enabled computations of strongly interacting dilute Fermi gases
with a high precision comparable to that of experimental measurements.
High-order analytic calculations that provide precision benchmarks
for QMC and experiment
represent an important tool for making further progress in this field. This is the focus of the present work.

Effective field theory provides the basis for such analytic benchmark calculations.
In this context, the problem of renormalization, which historically has presented a 
notable barrier for many-body calculations at high orders in perturbation theory, has been cleared up 
completely (in the perturbative case)~\cite{Polchinski:1983gv,Braaten:1996rq,Hammer:2000xg}.
While perturbative EFT calculations are generally restricted to low densities and weak interactions, respectively, 
they are still useful in many ways. 
Regarding the nuclear many-body problem~\cite{Bogner:2009bt,Holt:2013fwa,Hebeler:2015hla,Drischler:2019xuo,Drischler:2021kxf}, 
they provide viable input for constraining nuclear matter computations and neutron-star modeling.
Via resummation methods, they also give access to approximate analytic results of large-scattering length physics.

Here, we present in detail the calculation and results 
to fourth order in the perturbative EFT for zero-temperature many-fermion systems at very low energies, i.e.,  
the renowned Fermi-momentum or $k_{\rm F} a_s$ expansion 
for the ground-state energy of the dilute Fermi gas~\cite{Lenz,PhysRev.105.1119,1957PhRv,Efimov_1965,AMUSIA2,1965PhRv,Efimov2,AMUSIA1968377,RevModPhys.43.479,BISHOP1973106,PhysRevA.71.053605}.
In that, we follow up on our recent Letter~\cite{kfas1}
where the first fourth-order results have been presented.\footnote{We note the following typos in Ref.~\cite{kfas1}: in Eq.~(21) and (25) a factor $M^3$ is missing, and below Eq.~(24) it should read II6(ii) instead of III6(ii).}
In the present paper, we expand substantially on the results and presentation of Ref.~\cite{kfas1}.
First, in Sec.~\ref{chap2} we discuss in more detail the contact EFT formalism for fermions at very low energy scales.
In Sec.~\ref{chap3} we then present the details of the calculation of 
the Fermi-momentum expansion to fourth order for the case of spin one-half fermions.
The case of spins greater than one-half is examined in detail in Sec.~\ref{chap4} 
using two different regularization schemes: cutoff regularization and dimensional regularization.
Our fourth-order results 
for the ground-state energy of the general dilute Fermi gas are then summarized in Sec.~\ref{chap5}.
Using Bayesian methods, in Sec.~\ref{chap6} we investigate the convergence of the Fermi-momentum expansion.
There, we also study various Pad{\'e} and Borel approximants
constructed from the expansion.
Finally, Sec.~\ref{chap7} provides a short summary.

\section{Effective field theory for nonrelativistic fermions}\label{chap2}

The effective field theory (EFT) Lagrangian $\mathscr{L}_\text{EFT}$ for
dilute Fermi systems is composed of the most general two- and many-body contact interactions consistent
with Galilean invariance, parity, and time-reversal invariance.
Up to field redefinitions, its leading terms are given by 
(see, e.g., Refs.~\cite{Kaplan:1998we,Steele:2000qt,Hammer:2000xg,Furnstahl:2000we,Furnstahl:2001xq,Schafer:2005kg})
\begin{align} \label{Lagrangian} \nonumber
\mathscr{L}_\text{EFT} &= \psi^\dagger \bigg[i\partial_t +\frac{\ter \nabla^2}{2M} \bigg] \psi
-\frac{C_0}{2} (\psi^\dagger \psi)^2
\\ \nonumber
&\quad+\frac{C_2}{16} \left[(\psi \psi)^\dagger (\psi \te{\nabla}^2  \psi)+ \text{H.c.} \right]
\\ 
&\quad+\frac{C'_2}{8} (\psi \te{\nabla}\psi)^\dagger \cdot(\psi \te{\nabla}\psi)
- \frac{D_0}{6} (\psi^\dagger \psi)^3+ \ldots\,,
\end{align}
where $\psi$ are nonrelativistic fermion fields, ${\te{\nabla}=\tel \nabla-\ter \nabla}$ is the Galilean invariant derivative, H.c.~is the Hermitian conjugate, and $M$ is the fermion mass. The couplings of the contact interactions $C_0,C_2,C_2',D_0,\dotsc$, called
low-energy constants (LECs), have to be fit to experimental data or 
(if possible) matched to an underlying theory. (For recent work aimed at rooting contact EFT for nucleons
in lattice QCD calculations, see Refs.~\cite{Contessi:2017rww,Bansal:2017pwn}.)

A truncation scheme, known as power counting, is required to
organize the (infinite number of) EFT operators in a systematic way.
In particular, the power counting needs to renormalize the 
ultraviolet (UV) divergences at each order.
For perturbative calculations within the contact EFT given by Eq.~\eqref{Lagrangian}, the
power counting corresponds to
ordering
contributions in perturbation theory according to
the (naive) mass dimension $\sigma$ of the LECs, i.e., 
\begin{align} \label{NN-PC}
\sigma(C_{2n}^{(\prime)})&=2n+1\,,\\
\sigma(D_{2n}^{(\prime)})&=2n+4\,,\\
\sigma(E_{2n}^{(\prime)})&=2n+7\,,
\end{align}
etc., where the LECs $E_{2n}^{(\prime)}$ correspond to four-body interactions.

In the following, we first discuss in Sec.~\ref{chap20} the relation between
$N$-body scattering diagrams and the MBPT series for dilute Fermi systems.
This is followed by the analysis of UV power divergences and two-body scattering diagrams in Sec.~\ref{chap21}.
In Sec.~\ref{chap22} we then examine the ladder diagrams of MBPT.
Next, in Sec.~\ref{chap23} we study the renormalization of logarithmic UV divergences and the associated nonanalytic terms in the perturbative EFT expansion.
Finally, Sec.~\ref{chap24} briefly discusses different partial resummations for systems with a large $S$-wave scattering length.

\subsection{Renormalization from few-body to\newline many-body systems}\label{chap20}

The nonrelativistic field theory specified by the Lagrangian $\mathscr{L}_\text{EFT}$ is equivalent to a 
Hamiltonian approach with $N$-body potentials. The regularized two- and three-body potentials are given by
\begin{align} \nonumber
\braket{\bs{p'}|V^{(2)}_\text{EFT}|\bs{p}}
&= \Big[ C_0(\Lambda) + C_2(\Lambda) (\bs{p'}^2+\bs{p}^2)/2
\\ 
&\quad+ C_2'(\Lambda)\,\bs{p'}\cdot\bs{p}\,+ \ldots\Big]
 f(p/\Lambda)f(p'/\Lambda)\,,
\\ \nonumber
\braket{\bs{p'}\bs{q'}|V^{(3)}_\text{EFT}|\bs{p}\bs{q}}
&=
\Big[D_0(\Lambda)+\ldots\Big]  f(p/\Lambda)f(q/\Lambda)\\ 
&\quad \times
f(p'/\Lambda)f(q'/\Lambda)\,.
\end{align}
Here, $\bs{p^{(\prime)}}$ and $\bs{q}^{(\prime)}$ are relative and Jacobi momenta, respectively, and
$f(p/\Lambda)$ is a regulator function that suppresses high-momentum modes.
Later we will also consider dimensional regularization (DR), but for now we use a (Galilean invariant) momentum regulator.

The superficial degree of divergence $d$ of an $N$-body scattering diagram is given by
\begin{align} \label{dof}
d=5L-2I + \sum_{j=1}^\mathcal{V} [\sigma({g_j})-1]\,,
\end{align}
where $L$ is the loop number, $I$ the number of internal lines, $\mathcal{V}$ the number of vertices, and ${g_j \in \{C_{2n},D_{2n},\dots\}}$;
see, e.g., Refs.~\cite{Collins:1984xc,Drissi:2019lff} for details.
[If there are subdivergences the actual degree of divergence can be larger than $d$.]
The MBPT diagrams are obtained from scattering diagrams by closing the external lines (and excluding occupied states in loop integrals)
of a single scattering diagram, or by closing and connecting the external lines of several diagrams.
Since the hole propagators associated with closed external lines are bounded (or exponentially decaying at finite temperature),
the renormalization of MBPT follows from the renormalization of scattering diagrams.
For nonrelativistic contact interactions, $N$-body scattering diagrams can have only up to $N$ intermediate lines between adjacent vertices, so only $N'$-body interactions with ${N'\leqslant N}$ appear in a given the $N$-body sector.
This implies that the renormalization of the EFT interactions can be set up hierarchically, starting from
the renormalization of two-body interactions in the two-body sector, then three-body, and so on, up to a given truncation order in the power counting.

\subsection{Two-body scattering}\label{chap21}

\begin{figure*}[t]
\begin{center}
\includegraphics[trim= 0 25 0 0, width=0.9\textwidth]{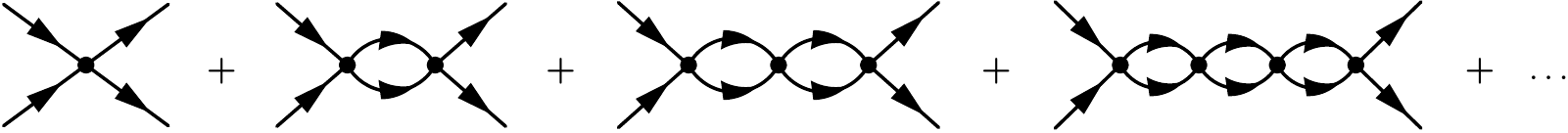}
\end{center}
\vspace*{0.4cm}
\caption{The two-body scattering diagrams. By closing the external lines one obtains the particle-particle ladder diagrams of MBPT.
The momentum integration associated with the closed lines has the effect 
that the (MBPT) ladder series has zero radius of convergence (renormalon divergence), in contrast to the 
series of two-body scattering diagrams (a geometric series). See the text for details.}
\label{fig_2to2}
\end{figure*}

In the nonrelativistic EFT, the only two-body scattering diagrams are ladder diagrams (corresponding
to iterations of the Lippmann-Schwinger equation), see Fig.~\ref{fig_2to2}.
This makes the two-body sector very simple: all loop integrals factorize, with factors $J_n(k,\Lambda)$ given by
\begin{align}
J_n(k,\Lambda)=
\int\limits_0^\infty \! d q \, \frac{q^{2n}}{k^2-q^2+i\epsilon}f^2(q/\Lambda)\,.
\end{align}
To extract the power divergence we rescale the loop momentum as $\bs{q}\rightarrow\bs{q}/\Lambda$, leading to
\begin{align}
J_n(k,\Lambda)=
I^\text{UV}_n(k,\Lambda)+J^R_n(k)\,,
\end{align}
where
\begin{align}
J^R_n(k)=\frac{i\pi}{2}k^{2n+1}
\end{align}
and $I^\text{UV}_n(k,\Lambda)=I^{\text{UV},\infty}_n(k,\Lambda)+I^{\text{UV},0}_n(k,\Lambda)$,
with
\begin{align}
I^{\text{UV},\infty}_n(k,\Lambda)&=-\sum_{m=0}^n \alpha_{2m}\Lambda^{2m+1} k^{2(n-m)}\,,
\\
I^{\text{UV},0}_n(k,\Lambda)&\xrightarrow{\Lambda\rightarrow \infty} 0\,,
\end{align}
where $\alpha_{2m}$ are regulator-dependent constants.
The effective-range expansion (ERE)
for the on-shell $T$ matrix reads~\cite{VANKOLCK1999273,Hammer:2000xg}
\begin{align} \label{ERE}
T(k,\cos\vartheta)&
= \frac{4\pi}{M}
\bigg\{
\underbrace{\sum_{n=0}^\infty \tau^{(s)}_n k^n}_{T^{(s)}(k)}
+
\underbrace{\sum_{n=2}^\infty \tau^{(p)}_n [k \cos\vartheta)]^n }_{T^{(p)}(k,\cos\vartheta)}
+
\ldots \bigg\}\,,
\end{align}
where $k$ and $\vartheta$ are the scattering momentum and angle, and
\begin{align}
\tau^{(s)}_n &=
\{
a_s,
- ia_s^2 ,
-a_s^3+a_s^2 r_s,
i(a_s^4-2 a_s^3r_s),
\breq
& \quad\quad
a_s^5-3 a_s^4r_s+a_s^3 r_s^2+a_s^2 v_s,
\ldots \}\,,
\\
\tau^{(p)}_n &=
\{
a_p^3,
\ldots \}\,,
\end{align}
with $a_s$ and $a_p$ the $S$- and $P$-wave scattering length, respectively, $r_s$ the $S$-wave effective range,
and $v_s$ the $S$-wave shape parameter.
Matching the regularized EFT perturbation series
to Eq.~\eqref{ERE} leads (in the infinite-cutoff limit ${\Lambda\rightarrow\infty}$) to
\begin{align} \label{couplings1}
C_0(\Lambda)
&= C_0 + C_0 \sum_{\nu= 1}^3 \left(\alpha_0C_0\frac{M}{2\pi^2}\Lambda\right)^{\!\nu}
+\alpha_2C_2 C_0 \frac{M}{3\pi^2} \Lambda^3
\breq & \quad
+ \ldots\,,
\\\label{couplings2}
C_2(\Lambda)
&= C_2
+\alpha_2C_2 C_0 \frac{M}{\pi^2}\Lambda + \ldots\,,
\\\label{couplings3}
C_2'(\Lambda)
&= C_2'+\ldots \,,
\end{align}
where
the cutoff-dependent parts
are counterterms that cancel UV divergences
and the omitted terms correspond to counterterms beyond fourth order.
Note that all the counterterms required to renormalize $C_0$-only contributions to the $T$ matrix are included in $C_0(\Lambda)$;
i.e., the $C_0$ term
corresponds to a perturbatively renormalizable interaction.
For spin multiplicities ${g>2}$, this
feature is however restricted to the two-body sector (see Sec.~\ref{chap23}).

The (renormalized) LECs are given by
\begin{align} \label{LECs}
C_0 = \frac{4\pi a_s}{M}, \hspace*{8mm} C_2=C_0\frac{a_sr_{\!s}}{2}, \hspace*{8mm} C_2'=\frac{4\pi a_p^3}{M}\,,
\end{align}
etc.
The perturbative EFT expansion is viable throughout the energy range appropriate to the EFT
only if the size of the LECs conforms to the power counting; i.e.,
\begin{align} \label{LECsScale}
C_0 \sim \frac{1}{M\Lambda_b}, \hspace*{9mm}
C_2 \sim C_2'\sim \frac{1}{M\Lambda_b^3}\,,
\end{align}
etc., 
corresponding to $a_s\sim r_s \sim a_p \sim 1/\Lambda_b$.
Here, $\Lambda_b$ is the ``hard scale'' beyond which the EFT description breaks down.
The scaling given by Eq.~\eqref{LECsScale}
is commonly referred to as the ``natural'' case~\cite{VANKOLCK1999273}.
The EFT perturbation series then corresponds to an expansion in powers of $Q/\Lambda_b$.

\subsection{Many-body ladder diagrams and renormalons}\label{chap22}

Closing the external lines of two-body scattering diagrams, one obtains the particle-particle (\textit{pp}) ladder diagrams of MBPT.
For these diagrams, the factors corresponding to the \textit{pp} bubbles are given by
\begin{align}
\mathscr{J}_n(P,k,\Lambda)=
\int\limits \! \frac{d^3q}{4\pi} \, \frac{q^{2n}}{k^2-q^2}\bar n_{|\bs{P}-\bs{q}|/2}\bar n_{|\bs{P}+\bs{q}|/2}f^2(q/\Lambda)\,,
\end{align}
where ${\bar n_k=\theta(k-\kF)}$, $\kF$ is the Fermi momentum,
and $\bs{q}$ is the relative momentum of the two particle lines in a given \textit{pp} bubble.
The hole lines correspond to integrating over $\bs{P}$ and $\bs{k}$.
The \textit{pp} bubble can be separated as
\begin{align}
\mathscr{J}_n(P,k)=
I^\text{UV}_n(k,\Lambda) + \mathscr{I}^R_n(P,k)\,,
\end{align}
where the cutoff-independent part is given by
\begin{align}
\mathscr{J}^R_n(P,k) &=
\int\limits \! \frac{d^3q}{4\pi} \, \frac{q^{2n}}{k^2-q^2}\left[\bar n_{|\bs{P}-\bs{q}|/2}\bar n_{|\bs{P}+\bs{q}|/2}-1\right]
\breq
&=
\frac{\kF}{2}+\frac{P}{2}+\frac{k}{2} \ln\left|\frac{\kF+P-k}{\kF+P+k}\right|
\breq & \quad +
\frac{\kF^2-P^2-k^2}{4P} \ln\left|\frac{(\kF+P)^2-k^2}{\kF^2-P^2-k^2}\right|\,.
\end{align}
Notably, 
the series of \textit{pp} ladder diagrams is a divergent asymptotic series with zero radius of 
convergence~\cite{Mari_o_2019,Feldman1996,RevModPhys.43.479}.
The physical context of this so-called ``renormalon divergence''
is the Cooper pairing phenomenon~\cite{Mari_o_2019}.
Mathematically, the divergence is due to the singularities of $\mathscr{J}^R_n(P,k)$ at the boundaries of the hole-line
integrals (i.e., the Lebesgue dominated convergence theorem is not 
satisfied).\footnote{Note that (in contrast to, e.g., relativistic $\phi^4$ theory~\cite{10.2307/j.ctt7zv6ds}) the renormalon divergence occurs for both the renormalized and the regularized perturbation series.
The MBPT series has still
zero radius convergence if the ladders are resummed~\cite{Mari_o_2019,PhysRevLett.121.130405}; the large-order behavior is however (expected to be)
dominated by renormalons~\cite{Mari_o_2019}.}

\begin{figure*}[t]
\begin{center}
\includegraphics[trim= 0 25 0 0, width=0.99\textwidth]{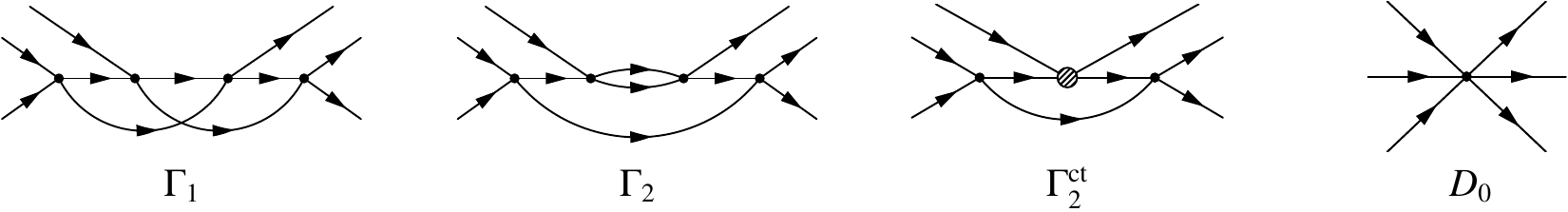}
\end{center}
\vspace*{0.4cm}
\caption{The first three-body scattering diagrams with logarithmic divergences $\Gamma_1$ and $\Gamma_2$. Also shown is the counterterm diagram
$\Gamma_2^\text{ct}$
for the \textit{pp} bubble of $\Gamma_2$ (the counterterm is depicted as a shaded blob). 
The fourth diagram is
the leading three-body contact contribution in three-fermion scattering, which 
includes the counterterm for the logarithmic UV divergences of $\Gamma_1$ and $\Gamma_2$.
Closing the external lines one obtains from $\Gamma_1$ and $\Gamma_2$ 
the MBPT diagrams with logarithmic divergences II5, II6, IIA1, and III1 shown in Fig.~\ref{MBPTdiags} below. 
See Sec.~\ref{chap4} for details on the evaluation of these diagrams.}
\label{fig_3to3}
\end{figure*}

\subsection{Multi-fermion scattering and logarithms}\label{chap23}

While the two-body scattering diagrams involve only UV power divergences [see Sec.~\ref{chap21}],
multi-fermion scattering involves also logarithmic divergences ${\sim \ln(\Lambda/Q)}$, where $Q$ is an invariant kinematical 
variable. For scattering diagrams, $Q$ is an external momentum, and in MBPT at zero temperature $Q$ is the Fermi momentum $\kF$.
That is, logarithmic UV divergences appear with a ratio of scales, which implies that their coefficients must be regulator independent
(in contrast to the coefficients of UV power divergences), see also Ref.~\cite{Hammer:2000xg}.
Renormalization removes the dependence on the UV cutoff $\Lambda$ such that the logarithms become $\ln(\Lambda_0/Q)$,
where
$\Lambda_0$ is an arbitrary auxiliary scale [see Sec.~\ref{chap4} for details].
The dependence on $\Lambda_0$ is canceled by the ``running'' with $\Lambda_0$ of the many-body coupling $g_j$ associated with the respective counterterm.
Note that this cancellation requires that the
involved terms are kept together, i.e., independent partial resummations are inhibited by the requirement of $\Lambda_0$ independence.

For ${g>2}$, the first logarithms in perturbative $N$-body scattering (for $N \geqslant 3$) appear
from the $C_0$ interaction
at order ${3N-5}$, i.e., at fourth order in the three-body sector.
(The first momentum-dependent logarithmic divergence appears at order
${3N-3}$ and is renormalized by $D_2$, etc., conforming to the perturbative EFT power counting.)
The fourth-order three-body scattering diagrams with logarithmic divergences, $\Gamma_1$ and $\Gamma_2$, are shown in Fig.~\ref{fig_3to3};
the associated many-body diagrams are listed in Fig.~\ref{MBPTdiags}.
They are renormalized by the contributions from the leading three-body contact interaction with coupling $D_0$ (corresponding to the last diagram in Fig.~\ref{fig_3to3}).  
This requires that the cutoff dependence of the $D_0$ coupling
is then given by 
\begin{align}
D_0(\Lambda)=D_0(\Lambda_0)+\eta M^3 C_0^4 \ln(\Lambda/\Lambda_0)\,.
\end{align}
The regulator-independent coefficient $\eta$ is obtained from the evaluation of the diagrams $\Gamma_1$ and $\Gamma_2$ (plus the 
bubble-counterterm diagram $\Gamma_2^\text{ct}$) of Fig.~\ref{fig_3to3}, 
or equivalently, 
from the evaluation of the corresponding many-body diagrams; see Sec.~\ref{chap4} for details.
The dependence of the first term $D_0(\Lambda_0)$ is such that $D_0(\Lambda)$ is independent of the auxiliary scale $\Lambda_0$. 
The value of $D_0(\Lambda_0)$ has to be fixed (for a given choice of $\Lambda_0$) 
by matching to few- or many-body data.

For ${g=2}$, all logarithmic
divergences from $S$-wave interactions cancel,
as required by the Pauli principle (the leading three-body contact interactions are Pauli blocked for ${g=2}$).
That is, for ${g=2}$ the $S$-wave part of the MBPT series is completely determined
by two-body scattering (i.e., by the ERE).
For $P$-wave interactions or $S$-wave interactions in ${g>2}$ systems on the other hand, 
an increasing number of $N$-body couplings is needed for perturbative renormalization beyond the two-body sector.

Finally, we note that the contact interactions between fermions can 
be rewritten such that they involve the propagation of (so-called) dimer fields, and
carrying out the partial diagrammatic resummations that renormalize the dimer propagator
makes the $C_0$ part of the perturbation series for three-body scattering
UV finite also for $g>2$~\cite{Bedaque:1998km}. 
Nevertheless, to achieve cutoff independence of the integral equation that corresponds to resumming
the remaining diagrams 
requires to include the leading three-body coupling $D_0$ (but no higher-order three-body interactions)~\cite{Bedaque:1998kg,Bedaque:1998km}. (Perturbatively expanding the nonperturbative three-body scattering amplitude then
allows one to determine the perturbative $D_0$ from nonperturbative three-body data~\cite{PhysRevLett.88.040401}.)
Beyond the two-body sector, the relation of the nonperturbative renormalization of the $C_0$ interaction (with a single three-body coupling $D_0$)
to the perturbative case, which requires in addition to $C_0$ and $D_0$ also many-body contact interactions at higher orders,
is thus nontrivial, and a general understanding of this issue
is still missing~\cite{Hammer:2019poc}.

\subsection{Resummations for large scattering length}\label{chap24}

If there is a two-body bound-state at threshold the $S$-wave scattering length $a_s$ is unnaturally large, 
and in this case the perturbative EFT expansion is of limited use.
In the two-body sector, this 
case can be straightforwardly dealt with by resumming the $C_0$ contributions and adding $C_2,\ldots$ perturbatively, which leads to~\cite{Kaplan:1998we,VANKOLCK1999273}
\begin{align} \label{ERE2}
T^{(s)}(k) &=
\frac{1}{\frac{1}{a_s}+ik}
+
\frac{r_s k^2}{\left(\frac{1}{a_s}+ik\right)^2}
+
\ldots \,.
\end{align}
Such a simple analytic resummation of the $C_0$ contributions is however not possible for the much more complicated MBPT series.
A notable benchmark for nonperturbative many-body treatments of the $C_0$ term 
is given by the ${a_s\rightarrow \infty}$ limit, corresponding to the unitary Fermi gas.
From dimensional analysis it follows that the ground-state energy density of the unitary Fermi gas of spin one-half fermions
is given by ${E(\kF)=\xi E_0(\kF)}$, where ${E_0(\kF)=\kF^5/(10\pi^2 M)}$ is the noninteracting ground-state energy density
and $\xi$ is the Bertsch parameter.
From experiments with ultracold atoms~\cite{Ku563}, the value ${\xi\approx 0.376(4)}$ has been inferred.

The most straightforward nonperturbative many-body approximation consists of resumming a subclass of MBPT diagrams.
The analytic resummation of the particle-particle (\textit{pp}) ladders gives ${\xi_\text{pp}\approx0.237}$~\cite{Schafer:2005kg,Kaiser:2011cg}, 
and resumming also hole-hole (\textit{hh}) and mixed \textit{pp}-\textit{hh} ladders gives 
${\xi_\text{ladders}\approx 0.5076}$~\cite{Kaiser:2011cg}.\footnote{The resummation of particle-hole ladders (``ring diagrams'') becomes relevant 
for large values of $g$, in particular regarding the expansion about the large-$g$ limit~\cite{Furnstahl:2002gt}.}
In addition, a value for ${\xi_\epsilon\approx 0.475}$ was 
deduced in Ref.~\cite{PhysRevLett.97.050403} by expanding in terms of ${\epsilon=4-d}$, where $d$ is the number of space dimensions, and subsequently interpolating 
between the ${\epsilon=2}$ and ${\epsilon=0}$ results for $E(\kF)$.
Even more close comes the value ${\xi_\text{LW}\approx 0.36}$ 
obtained from a self-consistent Luttinger-Ward type approach with resummed ladders~\cite{PhysRevA.75.023610} (see also Ref.~\cite{PhysRevA.95.043625}).
(See also Ref.~\cite{PhysRevLett.121.130405} for 
finite-temperature calculations based on Borel-resummed diagrammatic Monte Carlo calculations.)
The most accurate value has been obtained from QMC computations, 
${\xi_\text{QMC}= 0.372(5)}$~\cite{PhysRevA.84.061602}.

Predictions for $\xi$ may also be obtained 
by applying resummation methods such as Pad\'e approximants to the $a_s$ part of the Fermi-momentum expansion~\cite{Baker:1999np,baker4}. 
We will present results from this approach in Sec.~\ref{chap62}.

\section{Fourth-order term for\newline spin one-half fermions} \label{chap3}

We now start with the discussion of the perturbative EFT expansion for dilute many-fermion systems at fourth order. 
Logarithms and many-body interactions arise only for spin multiplicities ${g>2}$, the
intricacies of this case are postponed until Sec.~\ref{chap4}. Here, we discuss the spin one-half case, ${g=2}$, but 
we leave the notation general such that the results not affected by logarithmic terms can be carried over to ${g>2}$.
There are two different types of contributions at fourth order: 
(i) the second-order MBPT diagram with one $C_0$ and one $C_2$ vertex, 
and (ii) fourth-order MBPT diagrams with four $C_0$ vertices
(for ${g>2}$ there is also the first-order diagram with the $D_0$ vertex).
In each case, (in cutoff regularization) one has also two-body counterm contributions from lower-order MBPT diagrams.

\begin{figure*}[t]
\includegraphics[width=0.92\textwidth]{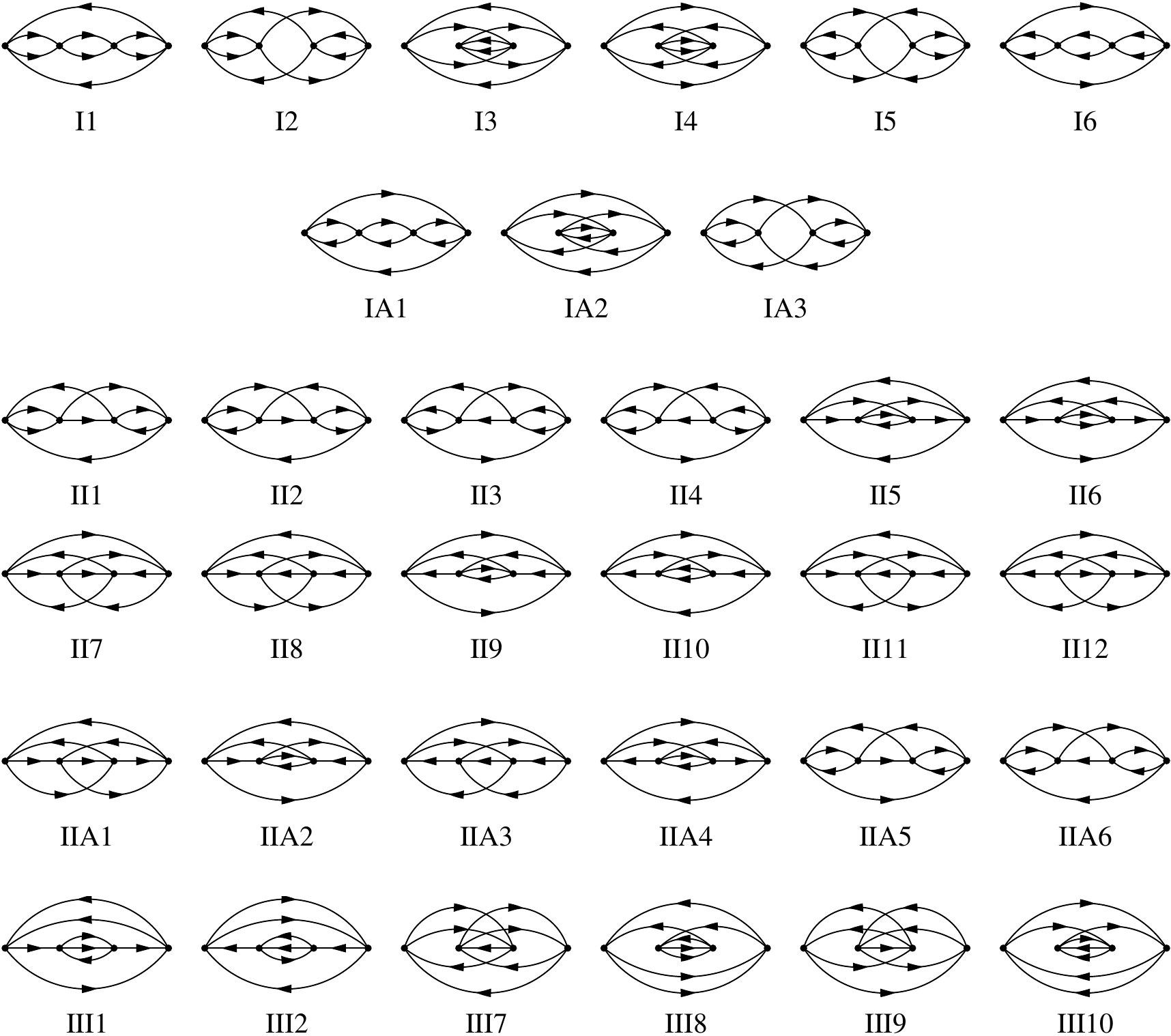} 
\caption{The 33 fourth-order Hugenholtz diagrams I(1-6), IA(1-3), II(1-12), IIA(1-6), and III(1,2,7-10.
Diagrams II5 and IIA1 (corresponding to $\Gamma_1$) as well as II6 and III1 (corresponding to $\Gamma_2$) have logarithmic UV divergences.}
\label{MBPTdiags}
\end{figure*}

For the calculation of the contribution (i), see Refs.~\cite{Kaiser:2012sr,kfas1}. 
The calculation of the contribution (ii)
is much more involved.
Among the possible MBPT diagrams with four $C_0$ vertices, only those without single-vertex loops have to be considered at zero temperature.
This is because all diagrams with single-vertex loops are removed by first-order mean-field (i.e., Hartree-Fock) insertions~\cite{PhysRevC.99.065811},
and for a momentum-independent interaction, first-order mean-field renormalization at zero temperature has no effect for a uniform system.
Therefore, as can easily be verified explicitly, these diagrams cancel each other at each order.
The 39 remaining fourth-order many-body diagrams can be divided into four topological species:
\begin{itemize}
\item I(1-6): ladder diagrams,
\item IA(1-3): ring diagrams,
\item II(1-12), IIA(1-6): other two-particle irreducible diagrams,
\item III(1-12): two-particle reducible diagrams,
\end{itemize}
where we have labeled diagrams according to groups that are
closed under permutations of the vertices: I(1–6), IA(1–3), II(1–12), IIA(1–6), III(1–12).
Diagrams III(3,6,11,12) are anomalous and thus give no contribution in zero-temperature MBPT~\cite{PhysRevC.99.065811,Kohn:1960zz}.
The 33 remaining diagrams are shown in Fig.~\ref{MBPTdiags}.

Diagrams I1, I6, and IA1 are the fourth-order versions of the third-order \textit{pp}, \textit{hh}, and \textit{ph}
diagrams; see, e.g., Ref.~\cite{Kaiser:2017xie}.
Diagrams I(2-5) are mixed \textit{pp}-\textit{hh} ladder diagrams.
The diagrams in the pairs I(3,4), III(7,8) and III(9,10) can be combined to get simplified energy denominators;
I(2,5), II(1,2), II(3,4), II(7,8), II(11,12) and IIA(2,4) give identical results for a spin-independent potential; and
for a momentum-independent potential the contribution from I(3+4) is half of that from I(2+5).

The ladder diagrams I(1-6) are most conveniently computed by expanding the semianalytic formula for the
ladder resummation derived by Kaiser~\cite{Kaiser:2011cg}.
The expressions obtained in this way can be derived from the usual many-body expressions by introducing relative momentum coordinates and applying various partial-fraction decompositions as well as
the Poincar\'{e}-Bertrand transformation formula~\cite{Muskheli}.
For the numerical evaluation of the IA diagrams, it is more convenient to use single-particle momenta instead of relative momenta, because then the
phase space is less complicated.
The II, IIA and III diagrams without divergences can be evaluated in the same way as the IA diagrams.
The following diagrams involve divergences:
\begin{itemize}
\item I(1,2,4,5), II(1,2,6), III(1,8): UV power divergences,
\item II(5,6), IIA1, III1: logarithmic UV divergences,
\item III(1,2,8,10): energy-denominator divergences.
\end{itemize}
The UV power divergences, corresponding to \textit{pp} bubbles, are renormalized in terms of (low-order) diagrams with two-body counterterm vertices.
For ${g=2}$, the logarithmic UV divergences cancel in the sums II5+IIA1 and II6+III1.
Finally, the energy-denominator divergences
correspond to higher-order poles at the integration boundary; they
cancel in the sums III(1+8) and III(2+10).

The counterterms for power divergences can be implemented by performing subtractions in the bubble parts of the integrands. For example,
using a sharp cutoff, ${f(p/\Lambda)=\theta(\Lambda-p)}$, and
scaling all momenta by a factor $\kF$, the regularized expression for II(1+2) is given by
\begin{align}
E_\text{4,II(1+2)}(\Lambda)&=
- \zeta (g-3) \sum_{ \substack{\bs{i},\bs{j},\bs{k} \\ \bs{d}}  }
\theta_{\bs{cd}}\theta_{\bs{kc}}\theta_{\bs{je}}\theta_{\bs{de}}
\frac{n_{ijk} \bar n_{cde}}{\mathcal{D}_{cd,ij}\mathcal{D}_{de,ik}}
\breq & \quad \times
\sum_{\bs{a}} \theta_{\bs{ab}}
\frac{\bar n_{ab}}{\mathcal{D}_{ab,ij}}
\bigg|{\substack{ {\color{white}dummy} \\\bs{b}=\bs{i}+\bs{j}-\bs{a}\\ \bs{c}=\bs{i}+\bs{j}-\bs{d}\\  \bs{e}=\bs{i}+\bs{k}-\bs{d}}}\,.
\end{align}
Here, ${\sum_{\bs{i}}\equiv \int d^3 i/(2\pi)^3}$, the distribution functions are ${n_{ij\ldots}\equiv n_i n_j \cdots}$ and
${\bar n_{ab\ldots}\equiv \bar n_a \bar n_b \cdots}$, with
${n_i\equiv\theta(1-i)}$ and ${\bar n_a\equiv\theta(a-1)}$, and
the energy denominators are given by ${\mathcal{D}_{ab,ij}\equiv(a^2+b^2-i^2-j^2)/(2M)}$.
Moreover, ${\zeta=\kF^9  g(g-1)C_0^4}$,
and ${\theta_{\bs{ab}}\equiv \theta(\Lambda/\kF-|\bs{a}-\bs{b}|/2)}$.
The dependence of a given MBPT diagram on $g$ is obtained
by inserting
a factor 
${\delta_{\sigma_1,\sigma'_1}\delta_{\sigma_2,\sigma'_2}-\delta_{\sigma_1,\sigma'_2}\delta_{\sigma_2,\sigma'_1}}$ for each vertex and summing over the 
spins $\sigma^{(\prime)}_1$, $\sigma^{(\prime)}_2$ of the in- and outgoing 
lines.
(For $P$-wave interactions the factor is ${\delta_{\sigma_1,\sigma'_1}\delta_{\sigma_2,\sigma'_2}+\delta_{\sigma_1,\sigma'_2}\delta_{\sigma_2,\sigma'_1}}$.) 
For details on the diagrammatic rules, see, e.g., Ref.~\cite{Hammer:2000xg,szabo}.
The renormalized expression is
given by\footnote{Throughout the paper, we label the cutoff-independent renormalized expressions corresponding to UV divergent diagrams with a subscript ``R''.}
\begin{align}
E^\text{R}_\text{4,II(1+2)}&=
-\zeta (g-3) \sum_{ \substack{\bs{i},\bs{j},\bs{k} \\ \bs{a},\bs{d}}  }
\frac{n_{ijk} \bar n_{cde}}{\mathcal{D}_{cd,ij}\mathcal{D}_{de,ik}}
\breq &\quad \times
\left[\frac{\bar n_{ab}}{\mathcal{D}_{ab,ij}} - \frac{1}{\mathcal{D}_{aa,00}} \right]
\bigg|{\substack{ {\color{white}dummy} \\ \bs{c}=\bs{i}+\bs{j}-\bs{d}\\  \bs{e}=\bs{i}+\bs{k}-\bs{d}\\ \bs{b}=\bs{i}+\bs{j}-\bs{a} }}\,,
\end{align}
where the part $\sim 1/\mathcal{D}_{aa,00}$ corresponds to the counterterm contribution.
This expression can be further simplified such that only one unbounded integral appears, i.e.,
using
\begin{align}  \label{dec}
\sum_{\bs{a}}
\left[\frac{\bar n_{ab}}{\mathcal{D}_{ab,ij}}  - \frac{1}{\mathcal{D}_{aa,00}} \right]
=
-
\sum_{\bs{a}}
\frac{n_a+n_b-n_{ab}}{\mathcal{D}_{ab,ij}}
\end{align}
we find
\begin{align}
E^\text{R}_\text{4,II(1+2)}&=
2\zeta (g-3) \sum_{ \substack{\bs{i},\bs{j},\bs{k},\bs{a} \\ \bs{d}}  }
\frac{n_{ijka} \bar n_{cde}}{\mathcal{D}_{cd,ij}\mathcal{D}_{de,ik}}
\frac{\mathcal{P}}{\mathcal{D}_{ab,ij}}
\bigg|{\substack{ {\color{white}dummy} \\ \bs{c}=\bs{i}+\bs{j}-\bs{d}\\  \bs{e}=\bs{i}+\bs{k}-\bs{d}\\ \bs{b}=\bs{i}+\bs{j}-\bs{a} }}\,,
\end{align}
where $\mathcal{P}$ denote the Cauchy principal value.
The direct application of Eq.~\eqref{dec} is prohibited for II6 and III1, because in that case
the pertinent energy denominators involve additional particle momenta.
The regularized expression for II6 is given by
\begin{align}
E_\text{4,II6}(\Lambda)&=
- \zeta (g-3)  \sum_{ \substack{\bs{i},\bs{j},\bs{k} \\ \bs{a},\bs{c}}  }
\theta_{\bs{ab}}\theta_{\bs{ka}}\theta_{\bs{cd}}\theta_{\bs{je}}\theta_{\bs{be}}
\breq & \quad \times
\frac{n_{ijk} \bar n_{abcde}}{\mathcal{D}_{ab,ij}\mathcal{D}_{be,ik} \mathcal{D}_{bcd,ijk}}
\Bigg|{\substack{ {\color{white}dummy} \\\bs{b}=\bs{i}+\bs{j}-\bs{a}\\ \bs{d}=\bs{k}+\bs{a}-\bs{c} \\ \bs{e}=\bs{k}+\bs{a}-\bs{j}}}\,.
\end{align}
Substituting $\bs{K}=(\bs{i}+\bs{j})/2$, $\bs{p}=(\bs{i}-\bs{j})/2$, $\bs{z}=\bs{k}$, $\bs{A}=(\bs{a}-\bs{b})/2$,
and $\bs{Y}=(\bs{c}-\bs{d})/2$, and omitting redundant regulator functions, we have
\begin{align}
E_\text{4,II6}(\Lambda)&=- 8M^3\,\zeta(g-3)   \sum_{ \substack{\bs{K},\bs{p},\bs{z} \\ \bs{A},\bs{Y}}}n_{ijk}  \bar n_{abcde}
\, \theta_A  \theta_Y\,\frac{1 }{A^2-p^2}
\breq&\quad\times
\frac{1 }{(\bs{A}+\bs{p})\cdot(\bs{A}-\bs{K}+\bs{z})}
\frac{1 }{Y^2+\mathcal{R}}\,,
\end{align}
where ${\mathcal{R}=(3\bs{A}+\bs{K}-\bs{z})\cdot(\bs{A}-\bs{K}+\bs{z})/4-p^2}$.
The UV power subdivergence can now be separated via
\begin{align} \label{II6sep}
\frac{1 }{Y^2-p^2+\mathcal{R}} =
\underbrace{\frac{1}{Y^2}}_{\leadsto E_\text{4,II6(i)}}
-
\underbrace{ \frac{\mathcal{R}}{(Y^2+\mathcal{R}) Y^2}}_{\leadsto E_\text{4,II6(ii)}}\,.
\end{align}
For the UV power divergence of II6(i), the counterterm can be implemented analogous to Eq.~\eqref{dec}.
The second part II6(ii) is only logarithmically UV divergent.
For ${g=2}$, the logarithmic divergence is canceled if we add the III1 term, which requires (due to the energy-denominator divergence)
to add also III(7+8).
The regularized expression for III(1+7+8) is given by
\begin{align}
E_\text{4,III(1+7+8)}(\Lambda)&=
-\zeta (g-1) \sum_{ \substack{\bs{i},\bs{j},\bs{k} \\ \bs{a},\bs{c} }  }
\theta_{\bs{ab}}\theta_{\bs{ab}}
\frac{n_{ijk} \bar n_{abc}}{\mathcal{D}_{ab,ij}^2 }
\breq & \quad \times
\left( \theta_{\bs{ka}}\theta_{\bs{cd}}\frac{\bar n_d}
{\mathcal{D}_{bcd,ijk}}
\right.
\breq & \quad
\left.
- \theta_{\bs{cd'}} \frac{\bar n_{d'}}
{\mathcal{D}_{cd',ik}} \right)
\Bigg|{\substack{{\color{white}dummy} \\\bs{b}=\bs{i}+\bs{j}-\bs{a}\\ \bs{d}=\bs{k}+\bs{a}-\bs{c}\\ \bs{d'}=\bs{i}+\bs{k}-\bs{c}}}\,.
\end{align}
The energy-denominator divergence corresponds to ${\mathcal{D}_{ab,ij}=0}$, and in that case the two terms in the large parentheses cancel each other.
For III(1+8) also
the linear UV divergences are removed.\footnote{The counterterms for the power divergences of III1 and III8 would come from diagrams with single-vertex loops.}
For ${g=2}$, the contribution from the sum II6(ii)+III(1+7+8) is then given by
\begin{align}
E^\text{R}_\text{4,II6(ii)+II(1+7+8)}\big|_{g=2} =&
-8 M^3 \zeta \sum_{ \substack{\bs{K},\bs{p},\bs{z} \\ \bs{A},\bs{Y}}}
\frac{n_{ijk} \bar n_{abc}}{A^2-p^2} \times \mathcal{G}\,,
\end{align}
with
\begin{align}
\mathcal{G}
&=
\frac{1}{(\bs{A}+\bs{p})\cdot(\bs{A}-\bs{K}+\bs{z})}
\frac{\mathcal{R}}{(Y^2+\mathcal{R}) Y^2}
\breq & \quad +
\frac{1 }{A^2-p^2}
\left[
\frac{1 }{Y^2+\mathcal{R}} - \frac{1 }{Y^2+\mathcal{R}'} \right]\,,
\end{align}
where ${\mathcal{R}'=-(\bs{K}+\bs{p}-\bs{Z})^2/4}$.
Finally, the regularized expressions for II5 and IIA1 are given by
\begin{align}
E_\text{4,II5}(\Lambda)&=
- \zeta (g-3) \sum_{ \substack{\bs{i},\bs{j},\bs{k} \\ \bs{a},\bs{c}}  }
\theta_{\bs{ab}}\theta_{\bs{kb}}\theta_{\bs{cd}}\theta_{\bs{ad}}\theta_{\bs{ke}}\theta_{\bs{ce}}
\breq & \quad \times
\frac{n_{ijk} \bar n_{abcde}}{\mathcal{D}_{ab,ij}\mathcal{D}_{ce,ij} \mathcal{D}_{acd,ijk}}
\bigg|{\substack{ {\color{white}dummy} \\ \bs{b}=\bs{i}+\bs{j}-\bs{a}\\ \bs{d}=\bs{i}+\bs{j}+\bs{k}-\bs{a}-\bs{c} \\ \bs{e}=\bs{i}+\bs{j}-\bs{c}}}\,,
\\
E_\text{4,IIA1}(\Lambda)&=
- \zeta (3g-5)  \sum_{ \substack{\bs{i},\bs{j},\bs{k} \\ \bs{a},\bs{c}}  }
\theta_{\bs{ab}}\theta_{\bs{kb}}\theta_{\bs{cd}}\theta_{\bs{ad}}\theta_{\bs{je}}\theta_{\bs{ce}}
\breq & \quad \times
\frac{n_{ijk} \bar n_{abcde} }{\mathcal{D}_{ab,ij}\mathcal{D}_{ce,ik} \mathcal{D}_{acd,ijk}}
\bigg|{\substack{ {\color{white}dummy} \\\bs{b}=\bs{i}+\bs{j}-\bs{a}\\ \bs{d}=\bs{i}+\bs{j}+\bs{k}-\bs{a}-\bs{c} \\ \bs{e}=\bs{i}+\bs{k}-\bs{c}}}\,.
\end{align}
For ${g=2}$, the sum of these contribution is UV finite, and is given by
\begin{align}
E^\text{R}_\text{4,II5+IIA1}\big|_{g=2} =&
\zeta \sum_{ \substack{\bs{i},\bs{j},\bs{k} \\ \bs{a},\bs{c}}  }
\frac{n_{ijk} \bar n_{abcd}}{\mathcal{D}_{ab,ij}\mathcal{D}_{acd,ijk}}
\bigg|{\substack{ {\color{white}dummy} \\{\color{white}dummy} \\  \bs{b}=\bs{i}+\bs{j}-\bs{a}\\\bs{d}=\bs{i}+\bs{j}+\bs{k}-\bs{a}-\bs{c}}}
\breq & \quad \times
\left(
\frac{ \bar n_e}{\mathcal{D}_{ce,ij}}
\bigg|{\substack{ {\color{white}dummy} \\{\color{white}dummy} \\ \bs{e}=\bs{i}+\bs{j}-\bs{c}}}
\right.
\breq & \quad
\left.
-
\frac{ \bar n_{e'}}{\mathcal{D}_{ce',ik}}
\bigg|{\substack{ {\color{white}dummy} \\{\color{white}dummy} \\\bs{e}'=\bs{i}+\bs{k}-\bs{c}}}
\right)\,.
\end{align}
The contributions from II5+IIA1 as well as II6(ii)+II(1+7+8)
can of course also be evaluated by subtracting the individual logarithmic divergences,
i.e., by adding the respective (counterterm) parts of $D_0(\Lambda)$ (only the sum of these parts vanishes for ${g=2}$), see Sec.~\ref{chap4}.
We have however found that evaluating the sums II5+IIA1 and II6(ii)+II(1+7+8) provides better numerical precision (see Table~\ref{ourtable}).

\section{Fourth-order term for\newline higher spins} \label{chap4}

For ${g>2}$,
the logarithmic divergences of II6, IIA1, II5 and III(1+7+8) are canceled by the 
contribution from the first-order MBPT diagram with $D_0$ vertex. In cutoff regularization, 
this cancellation is tantamount to
\begin{align} \label{RGE}
\Lambda \frac{\partial}{\partial\Lambda} D_0(\Lambda) = \eta M^3 C_0^4 \,,
\end{align}
where the coefficient $\eta$ is determined by the logarithmic UV divergence.
This can be integrated as
\begin{align}
D_0(\Lambda)
=
D_0(\Lambda_0) + \eta M^3 C_0^4 \ln(\Lambda/\Lambda_0)\,,
\end{align}
where
$\Lambda_0$ is an arbitrary auxiliary scale, and $D_0(\Lambda)$ is independent of $\Lambda_0$, 
as evident from the running with $\Lambda_0$ according to Eq.~\eqref{RGE} of the integration constant $D_0(\Lambda_0)$: 
$D_0(\Lambda_0')=D_0(\Lambda_0)+\eta M^3 C_0^4 \ln(\Lambda_0'/\Lambda_0)$.
The value of $D_0(\Lambda_0)$ has to be fixed (for a given choice of $\Lambda_0$) 
by matching to few- or many-body data (see, e.g., Ref.~\cite{PhysRevLett.88.040401}). For further details we refer to
the general discussion of logarithmic divergences in EFT provided in Sec.~\ref{chap23}.

Below, we first show how the fourth-order term for ${g>2}$
is calculated in cutoff regularization,
and then discuss the calculation in dimensional regularization (DR).
The pendant of Eq.~\eqref{RGE} in DR is given by Eq.~\eqref{RGEmu} below.

\subsection{Cutoff regularization}\label{chap41}

The coefficient ${\eta=\eta_1+\eta_2}$ is determined by the logarithmic divergence of II6+IIA1+II5+III(1+7+8), or equivalently, by
the logarithmic divergence of the three-body scattering diagrams $\Gamma_1$ and $\Gamma_2$ (see Fig.~\ref{fig_3to3}).
Using a sharp cutoff, ${f(p/\Lambda)=\theta(\Lambda-p)}$, the regularized expression for diagram $\Gamma_1$ is given by
\begin{align} \label{D1}
\Gamma_1(\Lambda)  &=
- \frac{3^3 2^4}{3!}C_0^4 \mathcal{J}_1(\Lambda) \,,
\end{align}
with
\begin{align} \label{D1J1}
\mathcal{J}_1(\Lambda)  &=
\sum_{\bs{x}_1,\bs{x}_2,\bs{x}_3,\bs{l}_1,\bs{l}_2} \!\!
\theta_{\bs{x}_1\bs{l}_1}
\theta_{\bs{x}_1\bs{k}_1}
\theta_{\bs{x}_2\bs{l}_1}
\theta_{\bs{x}_2\bs{l}_2}
\theta_{\bs{x}_3\bs{l}_2}
\theta_{\bs{x}_3\bs{k}_3'}
\breq & \quad \times
\frac{1}{\mathcal{D}^{*}_{x_1l_1,k_2k_3}
\mathcal{D}^{*}_{x_2 l_1 l_2,k_1k_2k_3}
\mathcal{D}^{*}_{x_3l_2,k_1'k_2'}}
\breq & \quad \times
\delta_{\bs{x}_1\bs{l}_1,\bs{k}_2\bs{k}_3}
\delta_{\bs{x}_2\bs{l}_1\bs{l}_2,\bs{k}_1\bs{k}_2\bs{k}_3}
\delta_{\bs{x}_3\bs{l}_2,\bs{k}_1'\bs{k}_2'}\,,
\end{align}
where $\bs{k}_{1,2,3}$ and $\bs{k}_{1,2,3}'$ are
the three-momenta of the in- and outgoing particles, respectively, with
${\bs{k}_1+\bs{k}_2+\bs{k}_3=\bs{k}'_1+\bs{k}'_2+\bs{k}'_3}$,
and $\bs{x}_{1,2,3}$ and $\bs{l}_{1,2}$
are the loop momenta, and ${\D^{*}=\D-i \epsilon}$.
The factor $3^3$ comes from cyclic permutations of the initial and final lines, the factor $2^4$ is due to the number of equivalent contractions for a given choice of final and initial lines, and the factor $1/3!$ is due to final-state antisymmetrization.
Similarly, the regularized expression for the sum of diagrams $\Gamma_{2}$ and $\Gamma_2^\text{ct}$ of Fig.~\ref{fig_3to3} is given by
\begin{align}\label{D2}
[\Gamma_2+\Gamma_2^\text{ct}](\Lambda)  &=
- \frac{3^2 2^3}{3!}C_0^4 \mathcal{J}_2(\Lambda) \,,
\end{align}
with
\begin{align}\label{D2J2}
\mathcal{J}_2(\Lambda)  &=
\sum_{\bs{x}_1,\bs{x}_2,\bs{x}_3,\bs{l}_1,\bs{l}_2} \!\!
\theta_{\bs{x}_1\bs{l}_1}
\theta_{\bs{x}_1\bs{k}_1}
\theta_{\bs{x}_2\bs{l}_2}
\theta_{\bs{x}_3\bs{l}_2}
\theta_{\bs{x}_3\bs{k}_1'}
\breq & \quad \times
\frac{1}
{\D^{*}_{x_1l_1,k_1k_2}
\D^{*}_{x_3l_1,k_1'k_2'}}
\left[\frac{1}{\D^{*}_{x_2 l_1 l_2,k_1k_2k_3}}-\frac{1}{\D^{*}_{l_2l_2}} \right]
\breq & \quad \times
\delta_{\bs{x}_1\bs{l}_1,\bs{k}_2\bs{k}_3}
\delta_{\bs{x}_2\bs{l}_1\bs{l}_2,\bs{k}_1\bs{k}_2\bs{k}_3}
\delta_{\bs{x}_3\bs{l}_1,\bs{k}_1'\bs{k}_2'}\,,
\end{align}
where the term in squared brackets involves the counterterm for the two-particle bubble.
Overall, the logarithmic divergence is given by
\begin{align}
\Gamma_1(\Lambda)
&\xrightarrow{\Lambda\rightarrow \infty}
-\eta_{1} M^3 C_0^4 \ln(\Lambda)\,,
\\
[\Gamma_2+\Gamma_2^\text{ct}](\Lambda)
&\xrightarrow{\Lambda\rightarrow \infty}
-\eta_{2} M^3 C_0^4 \ln(\Lambda)\,.
\end{align}
To determine $\eta_{1,2}$ we can set all external momenta to zero, i.e.,
\begin{align}
\mathcal{J}_{1,2}(\Lambda) \xrightarrow{\bs{k}^{(\prime)}_{1,2,3}\rightarrow 0} M^3\mathcal{I}_{1,2}(\Lambda)\,.
\end{align}
Introducing relative momenta $\bs{q}_{1}$ and $\bs{q_2}$ such that
$\{\bs{l}_{1,2},\bs{x}_{1,2,3}\}=\{\bs{q}_1,(\bs{q}_1+2\bs{q}_2)/2,\bs{q}_1,(\bs{q}_1-2\bs{q}_2)/2,-\bs{q}_1\}$,
the integral $\mathcal{I}_2(\Lambda)$ is given by
\begin{align}
\mathcal{I}_2(\Lambda) &= \!\!
\int\limits_{\aleph}^{\Lambda}  \frac{d^3q_1}{(2\pi)^3}
\int\limits_{\aleph}^{\Lambda}  \frac{d^3q_2}{(2\pi)^3}    \,
\frac{1}{q_1^4}\left[\frac{1}{3q_1^2/4+q_2^2}-\frac{1}{q_2^2}\right]\,,
\end{align}
where the integral boundaries are with respect to the radial coordinates. To have an infrared finite expression we have, as
a formal intermediate step, introduced an arbitrary infrared cutoff
$\aleph$.
This integral can be expressed in terms of the inverse tangent integral ${\text{Ti}_2(x)=[\text{Li}_2(ix)-\text{Li}_2(ix)]/(2i)}$, with $\text{Li}_2(z)$ the complex
dilogarithm, i.e.,
\begin{align}
- \frac{3^3 2^4}{3!}\mathcal{I}_2(\Lambda) &=
-\frac{\sqrt{3}}{2^3\pi^4}
\text{Ti}_2\left(
\frac{2q_2}{\sqrt{3}q_1 }
\right)\bigg|_{\aleph,\aleph}^{\Lambda,\Lambda}\,.
\end{align}
Using ${\text{Ti}_2(x) = \text{Ti}_2(1/x) + (\pi/2) \text{sgn}(x) \ln x}$ as well as ${\text{Ti}_2(0)=0}$ we find (for $\aleph \to 0$)
\begin{align} \label{D2logdiv}
\eta_2 =  -\frac{3\sqrt{3}}{4\pi^3}\,.
\end{align}
For $\Gamma_1$, this method to extract logarithms is prohibited by
the $\theta_{\bs{x}_3\bs{l}_2}$ factor and a nontrivial angular integral.
This problem can be avoided in DR where loop integrals remain invariant under
translations of the integration variables.
As shown in Sec.~\ref{chap42}, one obtains
\begin{align} \label{D1logdiv}
\eta_1 = \frac{1}{\pi^2} \,.
\end{align}
From this, the renormalized fourth-order contribution to the ground-state energy is
\begin{align}    \label{E4logdiv}
E_4(\kF) &=
\chi \big[D_0(\Lambda_0)+\eta M^3 C_0^4 \ln(\kF/\Lambda_0)\big] 
\breq & \quad
+ \sum_{i} E_\text{4,i}^\text{R}  + \ldots\,,
\end{align}
where ${i\in\{\text{II5,IIA1,II6,III(1+7+8)}\}}$, and the ellipses refer to contributions from other fourth-order diagrams.
The factor $\chi$ corresponding to the first-order three-body
diagram is
\begin{align} \label{xifactor}
\chi =\frac{g(g-1)(g-2)}{6}\sum_{ijk}n_{ijk} = \alpha (g-2) \frac{M}{108 \pi^4 a_s^4}\,,
\end{align}
with $\alpha = n \eps_\F (\kF a_s)^4 (g-1)$, where $n=g\,  \kF^3 /(6 \pi^2)$ is the fermion number density and $\eps_\F=\kF^2/(2M)$ the noninteracting Fermi energy.
Finally, the terms $E_\text{4,i}^\text{R}$ are given by\footnote{See Eq.~\eqref{II6sep} for the 
splitting of II6 into a power-divergent part II6(i) and a logarithmically divergent part II6(ii).}
\begin{align}
E_\text{4,II5}^\text{R}
&=\lim_{\Lambda\to\infty} \left[ E_\text{4,II5}(\Lambda)
+ (g-3)L_1(\Lambda)\right]\,,
\\
E_\text{4,IIA1}^\text{R}
&=\lim_{\Lambda\to\infty} \left[ E_\text{4,IIA1}(\Lambda)
+ (3g-5)L_1(\Lambda) \right]\,,
\\
E_\text{4,II6}^\text{R}
&=\lim_{\Lambda\to\infty} \left[ E_\text{4,II6(i)}^\text{R}+ E_\text{4,II6(ii)}(\Lambda) \right.
\\ \nonumber 
&\quad\quad\quad\quad \left.+(g-3)L_2(\Lambda)\right]\,,
\\
E_\text{4,III(1+7+8)}^\text{R}
&=\lim_{\Lambda\to\infty} \left[ E_\text{4,III(1+7+8)}(\Lambda)
+(g-1)L_2(\Lambda)\right]\,.
\end{align}
Here, the terms $L_1(\Lambda)$ and $L_2(\Lambda)$ 
cancel the logarithmic parts of the respective many-body diagrams, ${\sim\ln(\Lambda/\kF)}$,
with ${4(g-2)L_1(\Lambda)}+{2(g-2)L_2(\Lambda)} = {\chi \eta M^3 C_0^4 \ln(\Lambda/\kF)}$ matching the form of 
the logarithm in Eq.~\eqref{E4logdiv}.
They are given by 
\begin{align}
L_1(\Lambda)&=\alpha\, \frac{16}{27\pi^2}\ln(\Lambda/\kF) \,,
\\
L_2(\Lambda)&=- \alpha\,   \frac{8\sqrt{3}}{9\pi^3}\ln(\Lambda/\kF)\,,
\end{align}
which matches (with different phase-space prefactors) the logarithmic parts of 
the three-body scattering integrals $\mathcal{J}_1(\Lambda)$ and $\mathcal{J}_2(\Lambda)$, respectively.
One finds that
\begin{align}
E_\text{4,II5}^\text{R}
&= \alpha(g-3)\times 0.0645(1),
\\
E_\text{4,IIA1}^\text{R}
&= \alpha(3g-5)\times 0.0647(1) \,,
\\
E_\text{4,II6}^\text{R}
&= -\alpha(g-3)\times 0.0265(2) \,,
\\
E_\text{4,III(1+7+8)}^\text{R}
&= -\alpha(g-1)\times 0.0513(2)\,.
\end{align}
The sum of the first and second two contributions is given by
\begin{align} \label{sumlogs1,R}
E_\text{4,II5+IIA1}^\text{R} &= \alpha \big[0.00018(1) + (g-2)\times 0.2586(4)  \big]\,,
\\  \label{sumlogs2,R}
E_\text{4,II6+III(1+7+8)}^\text{R} &=  \alpha \big[-0.0248(1) 
\breq & \quad
-(g-2)\times 0.0778(3)  \big]\,,
\end{align}
where in each case the leading term corresponds to the result obtained in the ${g=2}$ calculation of Sec.~\ref{chap3}.

\subsection{Dimensional regularization}\label{chap42}

The DR calculation of the logarithmic terms is 
similar to the calculation of the corresponding terms for bosonic systems carried out by Braaten and Nieto~\cite{Braaten1999}. 
In DR, the coefficient of the logarithm arising from diagram $\Gamma_1$ is determined by the integral
\begin{align}
\mathcal{I}_1^{D} &=  \mu^{2(3-D)}
\int\!\!\frac{d^D l_1}{(2\pi)^D} \int\!\!\frac{d^D l_2}{(2\pi)^D} \frac{1}{(l_1^2+\aleph^2)}
\breq & \quad \times
\frac{1}{(l_1^2+l_2^2+\bs{l}_1\cdot \bs{l}_2+\aleph^2)}\frac{1}{(l_2^2+\aleph^2)}\,.
\end{align}
Here, $\mu$ is a momentum scale introduced to maintain the correct mass dimension, and
the scale $\aleph$ serves to admit the use of Eq.~\eqref{DRmasterformula} below. [Note that this is a different scale from the $\aleph$ used in Sec.~\ref{chap41} within cutoff regularization.]
Introducing Feynman parameters we obtain
\begin{align}
\mathcal{I}_1^{D} &= \mu^{2(3-D)}
\int\!\!\frac{d^D l_1}{(2\pi)^D} \int\!\!\frac{d^D l_2}{(2\pi)^D}
\int\limits_0^1 \! dx\int\limits_0^{1-x} \!\! dy
\breq &\quad \times
\frac{2}{\left[l_1^2(x+y)+l_2^2(1-x)+\bs{l}_1\cdot \bs{l}_2y+\aleph^2 \right]^3}\,.
\end{align}
Shifting $\bs{l}_1\rightarrow \bs{l}_1-\bs{l}_2 y/(2x+2y)$ and rescaling the integration variables leads to
\begin{align}
\mathcal{I}_1^{D} &=
\mu^{2(3-D)}
\mathcal{F}_1^D
\int\!\!\frac{d^D l_1}{(2\pi)^D} \int\!\!\frac{d^D l_2}{(2\pi)^D}
\frac{2}{(l_1^2+l_2^2+\aleph^2)^3}\,,
\end{align}
with
\begin{align}
\mathcal{F}_1^D
&=
\int\limits_0^1 \! dx\int\limits_0^{1-x} \!\! dy \left[(x+y)(1-x)-\frac{y^2}{4} \right]^{-D/2}\,.
\end{align}
We can expand this in $\veps=D-3$, $\mathcal{F}_1^{D=3+\veps}=\mathcal{F}_1+\veps \mathcal{F}_1' + \mathcal{O}(\veps^2)$, with
\begin{align}
\mathcal{F}_1
&=
\int\limits_0^1 \! dx\int\limits_0^{1-x} \!\! dy \left[(x+y)(1-x)-\frac{y^2}{4} \right]^{-3/2}
= \frac{4 \pi}{3}\,,
\\
\mathcal{F}_1'
&=-
\int\limits_0^1 \! dx\int\limits_0^{1-x} \!\! dy
\frac{\ln \left(x+y)((1-x) -\frac{y^2}{4}\right)}{2 \left[(x+y)(1-x) -\frac{y^2}{4}\right]^{3/2}}
\approx 4.71849\,.
\end{align}
Applying the relation~\cite{Collins:1984xc}
\begin{align} \label{DRmasterformula}
\int d^D \!\!  q\, \frac{1}{(q^2+\aleph^2)^n} &=
\pi^{D/2}
\frac{1}{\aleph^{2n-D}}
\frac{\Gamma(n-D/2)}{\Gamma(n)}\,,
\end{align}
and analytically continuing to ${D=3+\veps}$, we then find
\begin{align}\label{D1logDR}
\mathcal{I}_1^{D=3+\veps}
&=
\frac{1}{48\pi^2}
\left[-\frac{1}{\veps} -2\ln(\aleph/\mu) + \zeta_1 +  \mathcal{O}(\veps) \right ]\,,
\end{align}
where ${\zeta_1=\ln(4\pi)-\gamma_E- 3\mathcal{F}_1'/(4\pi) \approx 0.827352}$,
with ${\gamma_E \approx 0.577216}$ the Euler-Mascheroni constant.
This agrees with the corresponding result for bosonic systems
derived by Braaten and Nieto~\cite{Braaten1999}.
Note that for ${D=3}$ the left side of Eq.~\eqref{DRmasterformula} is
UV divergent for ${n\geqslant -3/2}$, but
the right side is singular only for ${n=3/2}$.
This is the well-known feature that power divergences are automatically set to zero in DR.

Efimov~\cite{Efimov_1965} and Bishop~\cite{BISHOP1973106} extracted the leading logarithms
by introducing a cutoff $\Lambda$ on one of the loop momenta $\bs{l}_{1,2}$ only.
DR makes it clear why this method gives the correct result:
the analytic continuation ${D\rightarrow 3+\veps}$ can be performed for individual subintegrals individually.
For diagrams with subdivergences this procedure would in fact be required to obtain finite results.
This is the case for diagram $\Gamma_{2}$, where the divergent integral is
\begin{align}
\mathcal{I}_{2,a}^{D} &=
\mu^{3-D}
\int\!\!\frac{d^D l_1}{(2\pi)^D} \,
\frac{1}{(l_1^2+\aleph^2)}\frac{1}{(l_1^2+\aleph^2)}
\breq & \quad \times
\mu^{3-D}
\int\frac{d^D l_2}{(2\pi)^D} \, \frac{1}{(l_1^2+l_2^2+\bs{l}_1\cdot \bs{l}_2+\aleph^2)}\,.
\end{align}
Shifting ${\bs{l}_2\rightarrow \bs{l}_2-\bs{l}_1/2}$, performing the $\bs{l}_2$ integration, and analytic continuation to ${D = 3+\veps}$ leads to
\begin{align}
\mathcal{I}_{2,a}^{D} &= -\frac{\mu^{3-D}}{4\pi}
\int\!\!\frac{d^D l_1}{(2\pi)^D}
\frac{1}{(l_1^2+\aleph^2)}\frac{1}{(l_1^2+\aleph^2)}
\sqrt{\frac{3}{4}l_1^2+\aleph^2}\,,
\end{align}
which, for ${D=3}$, indeed diverges only logarithmically in the UV.\footnote{In particular, setting ${D=3}$ and ${\aleph=0}$, and
introducing an UV cutoff on $\bs{l}_1$ is equivalent to the calculations by Efimov~\cite{Efimov_1965} and Bishop~\cite{BISHOP1973106}.}
Note that also the IR divergence (for ${\aleph=0}$, ${D=3}$) of the $\bs{l}_2$ integrals has been eliminated.
However, to get to a form where we can apply Eq.~\eqref{DRmasterformula} we would proceed instead as
\begin{align}
\mathcal{I}_{2,a}^{D} &=
\mu^{2(3-D)}
\int\!\!\frac{d^D l_1}{(2\pi)^D} \int\!\!\frac{d^D l_2}{(2\pi)^D}
\int\limits_0^1 \! dx (1-x)
\breq &\quad \times
\frac{2}{\left[l_1^2+l_2^2 x+\bs{l}_1\cdot \bs{l}_2 x+\aleph^2 \right]^3}\,.
\end{align}
Shifting ${\bs{l}_1\rightarrow \bs{l}_1-\bs{l}_2 x/2}$ and rescaling the integration variables leads to
\begin{align}
\mathcal{I}_{2,a}^{D} &=
\mu^{2(3-D)}
\mathcal{F}_{2,a}^D
\int\!\!\frac{d^D l_1}{(2\pi)^D} \int\!\!\frac{d^D l_2}{(2\pi)^D}
\frac{2}{(l_1^2+l_2^2+\aleph^2)^3}\,,
\end{align}
with
\begin{align}
\mathcal{F}_{2,a}^D
=
\int\limits_0^1 \! dx (1-x) \left[x-\frac{x^2}{4}\right]^{-D/2},
\end{align}
which is singular for ${D\geqslant 2}$, obviously a manifestation of the subdivergence.
This singularity can be removed by adding the term
\begin{align}
\mathcal{I}_{2,b}^{D} &=
-\mu^{2(3-D)}\int\frac{d^D l_1}{(2\pi)^D}
\frac{1}{(l_1^2+\aleph^2)}
\frac{1}{(l_1^2+\aleph^2)}
\breq &\quad \times
\int\frac{d^D l_2}{(2\pi)^D} \frac{1}{(l_2^2+\aleph^2)}\,,
\end{align}
i.e.,
\begin{align}
\mathcal{I}_{2,b}^{D} &=
\mu^{2(3-D)}
\mathcal{F}_{2,b}^{D}
\int\!\!\frac{d^D l_1}{(2\pi)^D} \int\!\!\frac{d^D l_2}{(2\pi)^D}
\frac{2}{(l_1^2+l_2^2+\aleph^2)^3}\,,
\end{align}
where
\begin{align}
\mathcal{F}_{2,b}^{D}
=
-\int\limits_0^1 \! dx (1-x) \left[x-x^2\right]^{-D/2}
\end{align}
is also singular for ${D\geqslant 2}$, but ${\mathcal{F}_{2}^{D}=\mathcal{F}_{2,a}^{D}+\mathcal{F}_{2,b}^{D}}$ is finite.
Expanding ${\mathcal{F}_{2}^{D=3+\veps}=
\mathcal{F}_2+\veps\mathcal{F}_2'+\mathcal{O}(\veps)^2}$, where
\begin{align}
\mathcal{F}_2
&=
-2\sqrt{3},
\\
\mathcal{F}_2'
&=
-\frac{8 \pi }{3}-\sqrt{3} \big(\ln \left(4/3\right)-2\big)
\approx -5.41176\,,
\end{align}
we find for ${\mathcal{I}_{2}^{D}=\mathcal{I}_{2,a}^{D}+\mathcal{I}_{2,b}^{D}}$:
\begin{align} \label{D2logDR}
\mathcal{I}_2^{D}
&=
-\frac{\sqrt{3}}{32\pi^3}
\left[-\frac{1}{\veps} -2\ln(\aleph/\mu) + \zeta_2 + \mathcal{O}(\veps) \right ]\,,
\end{align}
with ${\zeta_2=\ln(4\pi)-\gamma_E+ \mathcal{F}_2'/(2\sqrt{3}) \approx 0.39157}$.
This again matches the corresponding result for bosons derived by Braaten and Nieto~\cite{Braaten1999}.
As required, the coefficient of $\ln\mu$ matches the one of $\ln\Lambda$ in the cutoff calculation, see Eq.~\eqref{D2logdiv}.

Subtracting in Eqs.~\eqref{D1logDR} and~\eqref{D2logDR} only the divergent
parts ${\sim 1/\veps}$ corresponds to minimal subtraction (MS).
The coupling $D_0$ is then fixed as ${D_0=D^\star_0(\mu)+\eta M^3 C_0^4/(2\veps)}$, where
the scaling of $D^\star_0(\mu)$ with $\mu$ is identical
to the scaling of $D_0(\Lambda_0)$ with $\Lambda$, i.e.,
instead of Eq.~\eqref{RGE} we have
\begin{align} \label{RGEmu}
\mu \frac{\partial}{\partial\mu} D^\star_0(\mu) = \eta M^3 C_0^4 \,.
\end{align}
The couplings $D^\star_0(\mu)$ and $D_0(\Lambda_0)$
are not identical for ${\mu=\Lambda_0}$, i.e.,
they differ in terms of a subtraction constant specific to the respective
regularization and subtraction procedure.
Instead of Eq.~\eqref{E4logdiv} we have
\begin{align}   
E_4(\kF) &=
\chi \big[D_0^\star(\mu_0)+ \eta M^3 C_0^4 \ln(\kF/\mu_0)\big] \nonumber \\
& \quad + \sum_{i} E_\text{4,i}^\text{MS}  + \ldots\,,
\end{align}
with\footnote{For II6 we separate again the power-divergent part II6(i), see Eq.~\eqref{II6sep}.}
\begin{align}
E_\text{4,II5}^\text{MS}
&=E_\text{4,II5}(\aleph)
+ (g-3)L^\star_1(\aleph)\,,
\\
E_\text{4,IIA1}^\text{MS}
&=E_\text{4,IIA1}(\aleph)
+ (3g-5)L^\star_1(\aleph)\,,
\\
E_\text{4,II6}^\text{MS}
&=E_\text{4,II6(i)}^\text{R}+ E_\text{4,II6(ii)}(\aleph)
+ (g-3)L^\star_2(\aleph)\,,
\\
E_\text{4,III(1+7+8)}^\text{MS}
&=E_\text{4,III(1+7+8)}(\aleph)
+ (g-1)L^\star_1(\aleph)\,,
\end{align}
where
$4L_1^\star(\aleph)+2L_2^\star(\aleph) = \chi \eta M^3 C_0^4 \ln(\aleph/\kF)$, with
\begin{align}
L^\star_1(\aleph)&=\alpha\, \frac{16}{27\pi^2} \left[\frac{\zeta_1}{2}+\ln(\aleph/\kF) \right] \,,
\\
L^\star_2(\aleph)&=- \alpha\,   \frac{8\sqrt{3}}{9\pi^3} \left[\frac{\zeta_2}{2}+\ln(\aleph/\kF) \right]\,,
\end{align}
and the terms $E_\text{4,i}(\aleph)$ are given by
subtracting from the respective integrands
their values with the denominators replaced by those corresponding to $\mathcal{I}_{1,2}^{D}$.
For example, the term $E_\text{4,II5}(\aleph)$ is given by
\begin{align}
E_\text{4,II5}(\aleph)
&=
- \zeta (g-3) \sum_{ \substack{\bs{i},\bs{j},\bs{k} \\ \bs{a},\bs{c}} } n_{ijk}
\breq & \quad \times
\left[
\frac{\bar n_{abcde}}{\mathcal{D}_{ab,ij}\mathcal{D}_{ce,ij} \mathcal{D}_{acd,ijk}}
\Bigg|{\substack{ {\color{white}dummy} \\\bs{b}=\bs{i}+\bs{j}-\bs{a}\\ \bs{d}=\bs{i}+\bs{j}+\bs{k}-\bs{a}-\bs{c} \\ \bs{e}=\bs{i}+\bs{j}-\bs{c}}}
\right.
\breq & \quad
\left.
-
\frac{1}{\mathcal{D}^\aleph_{ab}\mathcal{D}^\aleph_{ce} \mathcal{D}^\aleph_{acd} }
\Bigg|{\substack{ {\color{white}dummy} \\ \bs{b}=-\bs{a}\\ \bs{d}=-\bs{a}-\bs{c} \\ \bs{e}=-\bs{c}} }
\right]\,,
\end{align}
with ${\mathcal{D}^\aleph_{ab}=\mathcal{D}_{ab}+\aleph^2/(\kF^2 M)}$.
One finds
\begin{align}
E_\text{4,II5}^\text{MS}
&= -\alpha(g-3)\times 0.0500(1)\,,
\\
E_\text{4,IIA1}^\text{MS}
&= -\alpha(3g-5)\times 0.0498(1) \,,
\\
E_\text{4,II6}^\text{MS}
&= \alpha(g-3)\times 0.0664(2) \,,
\\
E_\text{4,III(1+7+8)}^\text{MS}
&= \alpha(g-1)\times 0.0416(2)\,.
\end{align}
The sums of the first two and the last two contributions are given by
\begin{align}
E_\text{4,II5+IIA1}^\text{MS} &= \alpha \big[0.00018(1) - (g-2)\times 0.1995(4)  \big],
\\
E_\text{4,II6+III(1+7+8)}^\text{MS} &=  \alpha \big[-0.0248(1) +(g-2)\times 0.1079(2)  \big].
\end{align}
From Eqs.~\eqref{sumlogs1,R} and \eqref{sumlogs2,R}, the relation between the ``MS'' values and the ``R'' ones
is given by
\begin{align} 
E_\text{4,II5+IIA1}^\text{MS}
&=
E_\text{4,II5+IIA1}^\text{R} -
\alpha(g-2)\times 0.4581(8),
\\ \nonumber
E_\text{4,II6+III(1+7+8)}^\text{MS}
&=
E_\text{4,II6+III(1+7+8)}^\text{R}\\
&\quad +
\alpha(g-2)\times 0.1857(8)\,.
\end{align}
As required, the difference between the ``MS'' values and the ``R'' values
vanishes for ${g=2}$, see Sec.~\ref{chap3}.

\section{Ground-state energy at\newline fourth order}\label{chap5}

Here, we summarize the results for the low-density expansion for the ground-state energy density $E(\kF)$ of the dilute Fermi gas.
The expansion reads
\begin{align} \label{kfexp0}
E(\kF) = n \, \eps_\F \bigg[ \frac{3}{5} + (g-1) \sum_{\nu=1}^\infty \mathcal{C}_{\nu}(\kF)  \bigg],
\end{align}
with ${n=g\,  \kF^3 /(6 \pi^2)}$ the fermion number density, ${\eps_\F=\kF^2/(2M)}$ the noninteracting Fermi energy, and $g$ the spin multiplicity.
The expansion coefficients up to fourth order are given by
\begin{align}\label{E1}
\mathcal{C}_1(\kF) &= \frac{2}{3\pi} \kF a_s,
\\\label{E2}
\mathcal{C}_2(\kF) &= \frac{4}{35 \pi^2} (11- 2\ln 2) (\kF a_s)^2,
\\
\mathcal{C}_3(\kF) &= \Big[0.0755732(0) +  0.0573879(0) \, (g-3) \Big](\kF a_s)^3
\breq&\quad
+\frac{1}{10\pi} (\kF a_s)^2 \kF r_{\!s} +  \frac{1}{5\pi}\frac{g+1}{g-1}(\kF a_p)^3 ,
\\
\mathcal{C}_\text{4}(\kF) &=
-0.0425(1)\,(\kF a_s)^4
\breq&\quad
+  0.0644872(0)\,  (\kF a_s)^3  \kF r_{\!s}
\breq
&\quad +\gamma_\text{4}(\kF)\,(g-2)\, (\kF a_s)^4\,.
\end{align}
The first two terms are
the only ones for which closed-form expressions are known;
these where first derived by Lenz~\cite{Lenz} in 1929 and Lee and Yang~\cite{PhysRev.105.1119} as well as de Dominicis and Martin~\cite{1957PhRv} in 1957, respectively.
The third-order term was first computed by
de Dominicis and Martin~\cite{1957PhRv} in 1957
for hard spheres with two isospin states, by
Amusia and Efimov~\cite{AMUSIA2} in 1965 for a single species of hard spheres, and then
by
Efimov~\cite{Efimov2} in 1966 for the general dilute Fermi gas.
It was also computed subsequently by various authors~\cite{AMUSIA1968377,RevModPhys.43.479,BISHOP1973106,Hammer:2000xg,Kaiser:2011cg,Kaiser:2012sr,Kaiser:2017xie}.
Initial studies of the fourth-order term for ${g=2}$
were performed by Baker in Refs.~\cite{1965PhRv,RevModPhys.43.479,Baker:1999np,baker4}, see also Ref.~\cite{kfas1} for a discussion of these.

\begin{table}
\vspace*{-2.4mm}
\caption
{Results for the contributions to the regular (i.e., nonlogarithmic) $a_s^4$ part of $\mathcal{C}_\text{4}(\kF)$. Diagrams with $^{*}$ ($^{**}$) have UV power (logarithmic) divergences, which are subtracted by the respective counterterm contributions.
Diagrams with $^{***}$ have energy-denominator singularities.
For the diagrams with logarithmic divergences, ``(R)'' denotes the result obtained
using a regulator function and subtracting only divergent terms, and ``(MS)'' the result corresponding to DR with minimal subtraction.
The uncertainty estimates take into account both the statistical Monte Carlo uncertainties and variations of the cutoff.
The $g$ factors are listed without the generic factor $g(g-1)$.
See Fig.~\ref{MBPTdiags} for the diagrams.}
\begin{center}
\begin{ruledtabular}
\begin{tabular}{llll}
\addlinespace
diagram  & $g$ factor & value \\
\hline
I1$^{*}$                               & $1$        & $+0.0383115(0)$ \\
I2$^{*}$+I3+I4$^{*}$+I5$^{*}$          & $1$        & $+0.0148549(0)$ \\
I6                                     & $1$        & $-0.0006851(0)$ \\
IA1                                    & $g(g-3)+4$ & $-0.003623(1)$ \\
IA2                                    & $g(g-3)+4$ & $-0.001672(1)$ \\
IA3                                    & $g(g-3)+4$ & $-0.003343(1)$ \\
II1$^{*}$+II2$^{*}$                    & $g-3$      & $+0.058359(1)$ \\
II3+II4                                & $g-3$      & $-0.003358(1)$ \\
II5$^{**}(\text{R})$                    & $g-3$      & $+0.0645(1)$ \\
II5$^{**}(\text{MS})$                   & $g-3$      & $-0.0500(1)$ \\
II6$^{**,*}(\text{R})$                  & $g-3$      & $-0.0265(2)$ \\
II6$^{**,*}(\text{MS})$                 & $g-3$      & $+0.0664(2) $ \\
II7+II12                               & $g-3$      & $+0.003923(1)$ \\
II8+II11                               & $g-3$      & $+0.007667(1)$ \\
II9                                    & $g-3$      & $-0.000981(1)$ \\
II10                                   & $g-3$      & $-0.000347(1)$ \\
IIA1$^{**}(\text{R})$                            & $3g-5$     & $+0.0647(1)$ \\
IIA1$^{**}(\text{MS})$                            & $3g-5$    & $-0.0498(1)$ \\
IIA2+IIA4                              & $3g-5$     & $+0.004122(1)$ \\
IIA3                                   & $3g-5$     & $-0.000461(1)$ \\
IIA5                                   & $3g-5$     & $+0.003542(1)$ \\
IIA6                                   & $3g-5$     & $+0.003331(1)$ \\
III1$^{* {**},**,*}(\text{R})$+III7+III8$^{*{**},*}$  & $g-1$      & $-0.0513(2)$ \\
III1$^{* {**},**,*}(\text{MS})$+III7+III8$^{*{**},*}$  & $g-1$      & +$0.0416(2) $ \\
III2$^{*{**}}$+III9+III10$^{*{**}}$        & $g-1$      & $+0.001650(1)$ \\
\hline
(II5+IIA1)$_{g=2}$                       & $1$      & $+0.00018(1)$ \\
(II6+III1+III7+III8)$^{*}_{g=2}$             & $1$  & $-0.0248(1)$ \\
\hline
$\sum_{\text{diagrams}, g=2}$ &  $1$ & $-0.0425(1)$
\end{tabular}%
\end{ruledtabular}
\end{center}
\label{ourtable}
\end{table}

Up to third order, only two-body (i.e., ERE) parameters appear and the expansion is a polynomial in the Fermi momentum $\kF$.
At higher orders ${N\geqslant 4}$, logarithmic terms ${\sim \kF^n \ln(\kF/\Lambda_0)}$ enter,
starting at ${N=4}$ for ${g>2}$; for ${g=2}$, no logarithms emerge from $S$-wave interactions (as a consequence of the Pauli exclusion principle).
The logarithms 
are accompanied by many-body couplings [at fourth order, the coupling $D_0(\Lambda_0)$]
whose dependence on the auxiliary scale $\Lambda_0$ is such 
that the Fermi-momentum expansion is independent of $\Lambda_0$. 
The many-body couplings are renormalization scheme dependent and have to be 
matched to few-body (or many-body) observables calculated in the same scheme.
Using a Galilean invariant regulator function and subtracting only divergent terms (``R'' scheme), the ${g>2}$ part $\gamma_\text{4}(\kF)$
of the fourth-order term takes the form\footnote{The logarithmic part of Eq.~\eqref{gamma4r}
was first derived by Efimov~\cite{Efimov_1965,Efimov2} and subsequently in Refs.~\cite{AMUSIA1968377,BISHOP1973106,Braaten:1996rq,Braaten1999}.
Note that in the literature~\cite{Efimov_1965,Efimov2,AMUSIA1968377,BISHOP1973106,Braaten:1996rq,Braaten1999,Hammer:2000xg,PhysRevLett.88.040401}
the arbitrary scale $\Lambda_0$ is usually set to $\Lambda_0=1/a_s$.}
\begin{align}  \label{gamma4r}
\gamma_\text{4}^\text{R}(\kF)&= \frac{ M D_0(\Lambda_0)}{108\pi^4 a_s^4} +0.2707(4) - \, 0.00864(2)\,(g-2)
\breq
&\quad+\frac{16}{27 \pi^3}\left(4\pi-3\sqrt{3}\right)  \,
\ln(\kF/\Lambda_0)\,.
\end{align}
On the other hand, using DR with minimal subtraction (``MS'' scheme) one obtains
\begin{align}  \label{gamma4ms}
\gamma_\text{4}^\text{MS}(\kF)&= \frac{ M D^\star_0(\Lambda_0)}{108\pi^4 a_s^4} -0.0017(4) - \, 0.00864(2)\,(g-2)
\breq
&\quad+\frac{16}{27 \pi^3}\left(4\pi-3\sqrt{3}\right)  \,
\ln(\kF/\Lambda_0)\,.
\end{align}
The scaling of $D_0(\Lambda_0)$ and $D^\star_0(\Lambda_0)$ with $\Lambda_0$ is identical, and determined by the
$\Lambda_0$ independence of $\gamma_\text{4}(\kF)$.
The values of $D_0(\Lambda_0)$ and $D^\star_0(\Lambda_0)$ differ by a subtraction constant, i.e.,
\begin{align} \label{D0subconst}
D^\star_0(\Lambda_0) =  D_0(\Lambda_0) - \frac{108\pi^4a_s^4}{M} \times 0.2724(8) \,.
\end{align}
Although the subtraction constant is arbitrary, it is
nevertheless pertinent to specify its value (i.e., to specify the renormalization scheme)
in order to predict many-body results from few-body data, or vice versa.

The individual diagrammatic contributions
to the $C_0^4$ part of the fourth-order term are listed in Table~\ref{ourtable}.
The computations have been carried out
using the Monte Carlo framework introduced in Ref.~\cite{Drischler:2017wtt} to evaluate high-order many-body diagrams, see also Ref.~\cite{kfas1}.
The results for the contributions that involve logarithmic divergences, II5, II6, IIA1, and III(1+7+8), have the largest numerical uncertainties.
For $g=2$, slightly more precise results can be given for II5+IIA1 and II6+III(1+7+8), because then no logarithmic divergences occur (see Sec.~\ref{chap3}).

\section{Convergence analysis and resummations}\label{chap6}

As discussed above, for spin one-half fermions (${g=2}$) the logarithmic terms from $S$-wave interactions 
cancel (by virtue of the Pauli principle). Logarithms still arise from $P$-wave interactions at higher orders, i.e., at a certain order $N_\text{log}$. 
The Fermi-momentum expansion for ${\X=E/E_0}$, truncated at an order ${N<N_\text{log}}$, is thus a polynomial in ${\delta=\kF a_s}$:
\begin{align} \label{eq:EE0plot}
\X_N(\delta)=
1+\sum_{\nu=1}^N \x_\nu \delta^\nu\,,
\end{align}
where ${E_0 = 3 n\kF^2/(10 M)}$ is the energy density of the free Fermi gas, and  
the expansion coefficients ${\x_\nu \equiv \x_\nu(a_s, r_s, a_p,\dotsc)}$ 
are completely determined by the ERE. 
In the following, we
analyze the convergence behavior of Eq.~\eqref{eq:EE0plot} for two different cases.
First, we examine the case where all ERE parameters beyond $a_s$ are zero, which we denote by LO. 
Here, the coefficients in the $\kF a_s$~expansion are given by 
\begin{align} \label{eq:LO} 
\{\x_\nu\} =
\bigg\{ &\frac{10}{9\pi},
\frac{44-8\ln 2}{21\pi^2},\breq
& 0.0303089(0), 
-0.07076(39),\ldots\bigg\}\,.
\end{align}
Second, we consider the hard-sphere gas (HS) where $a_s=3r_s/2=a_p$, leading to
\begin{align} \label{eq:HS} 
\{\x_\nu\} =
\left\{ \frac{10}{9\pi},
\frac{44-8\ln 2}{21\pi^2},
0.383987(0),
0.00089(39),\ldots\right\}\,.
\end{align}
In Sec.~\ref{chap61} 
we examine the convergence behavior of 
the LO and HS expansions and analyze the uncertainties of the predictions for $E/E_0$.
We will find that in both cases 
the Fermi-momentum expansion 
is well-converged at fourth-order for  $|\delta| \lesssim 0.5$.
In Sec.~\ref{chap62} we then show that Pad\'e and Borel resummations allow us to extend the  domain of convergence to $|\delta| \lesssim 1$.
Finally, in Sec.~\ref{chap63} we discuss the challenges regarding the calculation of 
the Fermi-momentum expansion beyond fourth order.

\subsection{Perturbative convergence and\newline uncertainty estimates}\label{chap61}

In Ref.~\cite{kfas1} we assessed the convergence pattern of the
$\kF a_s$~expansion at a given order ${N \leqslant 4}$ 
by setting the next-higher coefficient ${\x_{N+1} = \pm \max \left[ \x_{\nu\leqslant N} \right]}$. 
This spans an uncertainty band of width  ${\Delta \X_N=2 \left| \x_{N+1} \right|\delta^{N+1}}$. 
Here, we use the pointwise Bayesian model with
conjugate distributions developed in Refs.~\cite{Melendez:2019izc,Melendez:2017phj} to estimate $\x_{N+1}$ given the computed coefficients. This model allows one to evaluate posterior distributions
analytically (given the conjugate prior) rather than through Monte Carlo
sampling. Specifically, we treat the coefficients $\x_{\nu}$ as random numbers
drawn from a single normal distribution,\footnote{$z \sim \cdots$ is a common notation in statistics that reads ``the variable $z$ is
distributed as $\cdots$''. The ``\iid'' above the $\sim$ indicates a set of independent and identically distributed (\iid) random variables.}
\begin{align}
  \pr \left( \x_{\nu} \given \cbar^2 \right) \overset{\iid}{\sim} 
  \frac{1}{\sqrt{2\pi\cbar^2}}\exp\left[-\frac{\x_{\nu}^2}{2\cbar^2}\right]\,,
\end{align}
with mean zero and variance $\cbar^2$. 
The computed coefficients $\x_{\nu \leqslant 4}$ are assumed to be known draws from this \emph{a~priori} unknown distribution function, while $\x_{\nu>4}$ are unknown.
We also assume a scaled inverse-$\chi^2$ prior on~$\cbar^2$,
\begin{align}
  \pr\left( \cbar^2 \right) \sim 
  \frac{(\tau_0^2\frac{\eta_0}{2})^{\frac{\eta_0}{2}}}{\Gamma(\frac{\eta_0}{2})}~\frac{\exp\left[ -\frac{\eta_0 \tau_0^2}{2 \cbar^2}\right]}{\cbar^{2(1+\frac{\eta_0}{2})}}\,,
\end{align}
with $\eta_0$ degrees of freedom and scale parameter $\tau_0$. By adjusting
the hyperparameters we can incorporate our prior estimate of the (not computed)
higher-order coefficients. 
We fix ${\eta_0 = 3}$ and determine $\tau_0^2$ by the requirement that the mean
value $\eta_0 \tau_0^2/(\eta_0 - 2)$ equals ${\left| \max \left
[\x_{\nu\leqslant N} \right] \right|}$.
This prior choice disfavors high values for $\cbar^2 $ and thus $\x_{\nu>N}$. Using
Bayes' theorem and marginalizing over $\cbar^2$, one then finds that the posterior for a coefficient at
order ${n > N}$ is given by the Student's $t$ distribution~\cite{Melendez:2019izc}, i.e.,
\begin{align}\label{student-t}
    \pr \left( \x_{\nu>N}  \given \{ \x_\nu\}_{\nu = 1}^{N}, \tau^2 \right) \sim t_\eta \left( \x_\nu; 0, \tau^2 \right)\,,
\end{align}
with
\begin{align}
t_\eta(x; \mu,\tau^2)=
\frac{1}{\sqrt{\pi\eta\tau^2}} 
\frac{\Gamma(\frac{\eta+1}{2})}{\Gamma(\frac{\eta}{2})}
\left(1+\frac{(x-\mu)^2}{\eta\tau^2}\right)^{-\frac{\eta+1}{2}}\,.
\end{align}
Here, the scale parameter $\tau^2$ satisfies
\begin{align}
  \eta \tau^2 = \eta_0 \tau_0^2 + \sum_{\nu = 1}^N \x_\nu^2 \,.
\end{align}
Furthermore, ${\eta = \eta_0 + n_c}$, where ${n_c}$ is the number of coefficients in the set $\{ \x_\nu\}_{\nu = 1}^{n_c}$ 
used to inform the probability distribution.
We consider all available
coefficients, i.e., $n_c = 4$, so that all four known coefficients are used
for each $N\in\{1,2,3,4\}$ in Eq.~\eqref{student-t}.
Finally, from Bayes' theorem one then finds that the posterior distribution 
representing the uncertainty of 
$\X_N(\delta)$ is given by~\cite{Melendez:2019izc}
\begin{align}
  \pr \left( \X_N(\delta) \given  \{ \x_\nu\}_{\nu = 1}^{N},\tau^2 \right) \sim t_\eta \left( \X(\delta); \X_N(\delta), \delta^{2(N+1)} \tau^2 \right)\,,
\end{align}
where the variable $\X(\delta)$ corresponds to the presumed exact results.

\begin{figure*}[t]
\begin{center}
\vspace*{-4mm}
\includegraphics[width=\textwidth, page=5]{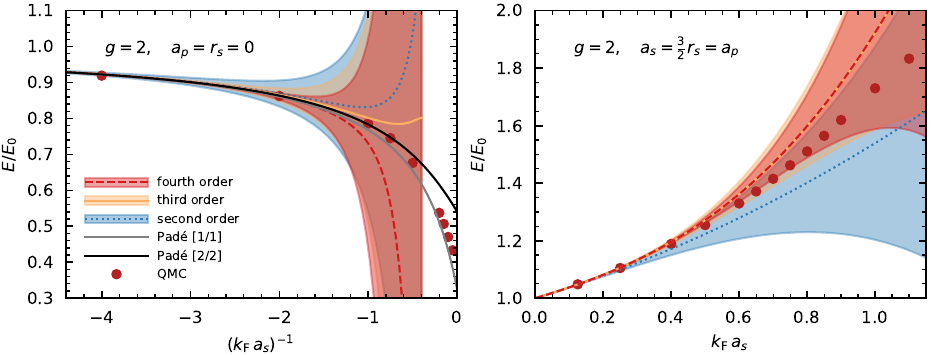}
\end{center}
\vspace*{-4mm}
\caption{
Convergence behavior of the Fermi-momentum expansion for the ground-state energy $E/E_0$ 
of a dilute Fermi gas of spin one-half fermions with $a_p=r_s=0$ (LO, left panel)
and $a_s=3r_s/2=a_p$ (HS, right panel) at negative and positive $\kF a_s$, respectively.  
The respective uncertainty bands correspond the the 68\% credibility intervals from
our Bayesian estimation of the next-higher coefficient in the $\kF a_s$ expansion.
In the LO case we also show results obtained from two different Pad{\'e} approximants, Pad{\'e}$[1,1]$ (gray line) and Pad{\'e}$[2,2]$
(black line).
Finally, the thick red dots in each panel correspond to results from nonperturbative QMC
computations~\cite{Gandolfi:2015jma,Gandolfi,Pila10FerroQMC}. 
Note that the $x$-axes in the two panels are different, and 
based on the available QMC data we show the attractive regime with $a_s<0$ in the left 
panel and the repulsive regime with $a_s>0$ in right panel.
See the text for more details.}
\label{fig:plot}
\end{figure*}

The convergence behavior of the Fermi-momentum expansion for the LO and the HS case is examined in Fig.~\ref{fig:plot}.
There, we show the perturbative results for $\X=E/E_0$ obtained for truncation orders $N=2,3,4$ together with the respective
68\% credibility intervals of our Bayesian analysis.
Also shown are data points
obtained from nonperturbative QMC
computations~\cite{Gandolfi:2015jma,Gandolfi,Pila10FerroQMC}.
One sees that the perturbative results are very close to the QMC data for
${|\delta| \lesssim 0.5}$ but start to deviate strongly for $|\delta| \gtrsim 1$.
In the LO case the relative error with respect to the QMC point at
${\delta = - 0.5}$ (${\X_\text{QMC} \approx 0.862}$) is $4.5\%$ (${\X_1 \approx 0.823}$) at first, 
$0.8\%$ (${\X_2 \approx 0.870}$) at second, $0.4\%$ (${\X_3 \approx 0.866}$) at third, and
$0.1\%$ (${\X_4 \approx 0.861}$) at fourth order, while 
in the HS case the relative error at
${\delta = + 0.5}$ (${\X_\text{QMC} \approx 1.254}$) is $6.2\%$ (${\X_1 \approx 1.177}$) at first, 
$2.5\%$ (${\X_2 \approx 1.223}$) at second, $1.3\%$ (${\X_3 \approx 1.271}$) at third, and
$1.3\%$ (${\X_4 \approx 1.271}$) at fourth order. 
The convergence of the expansion is slower in the HS case, which is signified by the relatively large size of the third-order coefficient there, 
${\x_3\approx 0.38}$ (in the LO case it is ${\x_3\approx 0.03}$).
The fourth-order HS coefficient ${\x_4\approx 0.0009}$ on the other hand is very small (due to a large cancellation between $S$- and $P$-wave contributions), so the
third- and fourth-order HS curves in Fig.~\ref{fig:plot} are almost indistinguishable.

The Bayesian uncertainty bands in Fig.~\ref{fig:plot}
are similar to those from the simple ${\x_{N+1} = \pm \max \left[ \x_{\nu\leqslant N} \right]}$ analysis, see Fig.~2 of Ref.~\cite{kfas1}.
In both schemes, going to higher
orders in the expansion reduces the width of the uncertainty bands for $|\delta| \lesssim 1$, and
for ${|\delta| \lesssim 0.5}$ the bands
are very small for $N=4$. 
This supports the conclusion that the expansion is
well-converged at fourth order for $|\delta| \lesssim 0.5$, 
and diverges for $|\delta| \gtrsim 1$.\footnote{More precisely, 
the Fermi-momentum expansion is an asymptotic series that diverges for $N\rightarrow \infty$ for all $|\delta|>0$, 
see Sec.~\ref{chap22}. 
``Well-converged'' means here that the result seems to be insensitive to lowering the truncation order.
}
Note that these results do not
depend on $a_s$ being of natural size; only $\kF a_s$ has to be small.

\subsection{Pad\'e and Borel resummations}\label{chap62}

Resummation methods provide a means to extrapolate a (truncated) series beyond the region where well-converged results are 
obtained, ${|\delta| \lesssim 0.5}$ in the present case.  
The two most common methods are Pad\'e approximants~\cite{bakerbook,benderbook} 
and Borel resummation~\cite{ZinnJustin:1980uk,Kleinert:2001ax,DORIGONI2019167914,ANICETO20191}.
Below, we apply these two methods to the Fermi-momentum expansion for the LO case (with negative $\delta$).
We do not consider the HS case, because
higher-order ERE parameters become relevant there at stronger coupling.
Regarding Pad\'e approximants, we restrict the discussion to those that give predictions for the Bertsch parameter 
${\xi=\X(-\infty)}$.\footnote{Pad\'e predictions for the Bertsch parameter were previously 
studied by Baker~\cite{Baker:1999np,baker4}, see also Ref.~\cite{kfas1}.}
In the Borel case we focus 
on the region of weak-to-intermediate coupling since only there (i.e., for ${|\delta| \lesssim 1}$) the 
extrapolations are well converged.

\subsubsection{Pad\'e approximants}

For a
given formal power series 
\begin{align} \label{formalseries}
\X(\delta)= 1+\sum_{\nu=1}^\infty \x_\nu \delta^\nu\,,
\end{align}
the
Pad{\'e}$[n,m]$ approximant is the rational function
\begin{align}
\text{Pad{\'e}}[n,m](\delta)=1+\frac{\sum_{k=1}^n {a_k} \delta^k}{1+\sum_{l=1}^m b_l \delta^l} \,,
\end{align}
whose Maclaurin expansion matches the series up to order ${N=n+m}$. Only ``diagonal'' Pad\'es
with ${n = m}$ have a nontrivial unitary limit, i.e., ${\text{Pad{\'e}}[n,n] \longrightarrow 1+a_n/b_n}$ for ${\delta \to -\infty}$. 
To have meaningful results in the strong-coupling regime thus 
mandates the restriction to even ${N=2n}$, i.e., ${(N,n)=(2,1)}$ and ${(N,n)=(4,2)}$.

The results obtained from the Pad{\'e}$[1,1]$ and $[2,2]$ approximants (which were already studied in Ref.~\cite{kfas1})
are shown in the left panel of Fig.~\ref{fig:plot}.\footnote{For a more extensive study of 
the Pad{\'e}$[1,1]$ approximant, see Ref.~\cite{Boulet:2019wfd}.}
One sees that the Pad{\'e}$[2,2]$ approximant is very close to the QMC results for 
${\delta\lesssim -1.2}$, while Pad{\'e}$[1,1]$ is in better agreement close to the unitary
limit ${\delta\rightarrow -\infty}$. Note however that pairing correlations become
relevant for larger values of $-\delta$, and it is questionable that Pad{\'e}s can 
capture pairing effects (which are expected to be encoded in the
high-order behavior of the $\kF a_s$ expansion~\cite{Mari_o_2019}) at low truncation orders. 
The range for the
Bertsch parameter obtained from Pad{\'e}$[1,1]$ and $[2,2]$,
${\xi_\text{Pad{\'e}}\in[0.326,0.541]}$, is consistent with the value 
${\xi\approx0.376}$ extracted from experiments with cold atomic gases, and also with the
extrapolated value for the normal (i.e., nonsuperfluid) Bertsch parameter
${\xi_n\approx 0.45}$~\cite{Ku563}. Altogether, these results seem to indicate
that Pad{\'e} approximants converge in a larger region, compared to the Fermi-momentum
expansion.

\begin{figure}[t]
\begin{center}
\vspace*{-4mm}
\includegraphics[width=0.48\textwidth]{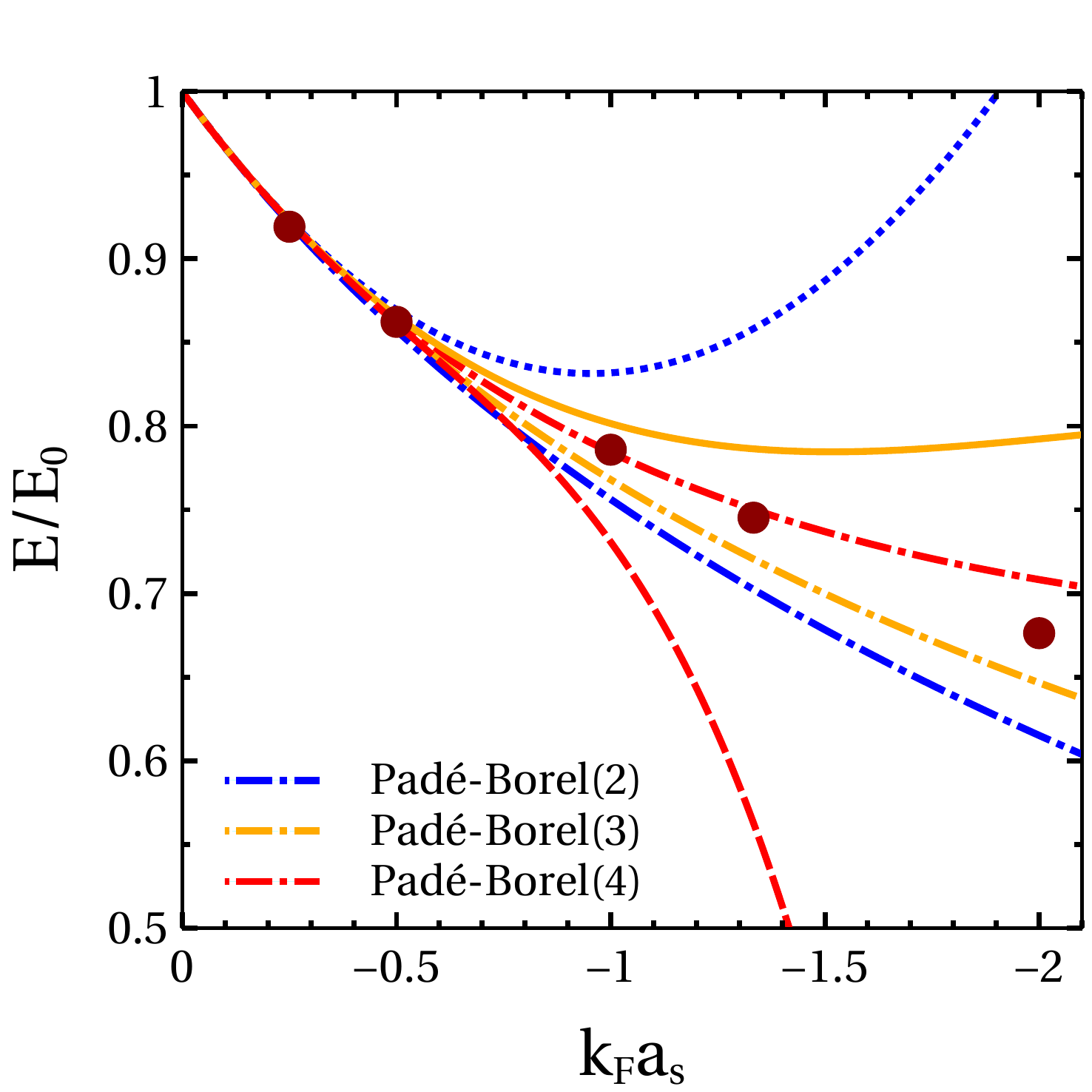}
\end{center}
\vspace*{-4mm}
\caption{
Results for the
ground-state energy $E/E_0$ 
of a dilute Fermi gas of spin one-half fermions with ${a_p=r_s=0}$
obtained from the Pad{\'e}-Borel resummation method, see the text for details.
The numbers in parentheses denote the underlying truncation order $N$.
Also shown are results from QMC
computations (filled red circles) as well as the
perturbative results at second (dotted blue line), third (solid yellow line) and fourth order (dashed red line), see also Fig.~\ref{fig:plot}.}
\label{fig:borel}
\end{figure}

\subsubsection{Borel resummation}

Borel resummation is based on the Borel(-Leroy) transformed perturbation series, i.e.,
\begin{align} \label{Borelseries}
\mathcal{B}(t)
\simeq
1+\sum_{\nu=1}^\infty \frac{\x_\nu}{\Gamma(\nu+1+\beta_0)}t^\nu\,,
\end{align}
where the standard Borel transform corresponds to ${\beta_0=0}$.
In contrast to the perturbative series [Eq.~\eqref{formalseries}], the Borel transformed series has a finite convergence radius:
from the
large-order behavior 
\begin{align} \label{PTlargeorder}
\x_\nu\stackrel{\nu\rightarrow \infty}{\sim} a^\nu\Gamma(\nu+1+\beta),
\end{align} 
one finds that 
the leading singularity of $\mathcal{B}(t)$
is at ${t=1/a}$~\cite{Kleinert:2001ax,Costin_2019,PhysRevD.100.056019}.
Formally, in the so-called Borel-summable case where all singularities of $\mathcal{B}(t)$ are off the positive real axis (in particular, $a<0$),
the exact $\X(\delta)$ is then obtained by first analytically continuing $\mathcal{B}(t)$ beyond ${t=1/|a|}$ 
and then carrying out the inverse Borel transform:
\begin{align}\label{Laplacetraf}
\X(\delta) =\int_{0}^\infty  \!\! dt \e^{-t} t^{\,\beta_0} \mathcal{B}(t\delta)\,. 
\end{align}
Regarding practical applications where the perturbative series is only 
known up to a finite order $N$, 
this procedure allows one to construct approximants $B_N(\delta)$ for $\X(\delta)$
in terms of approximants $\mathcal{B}_N(t)$ for $\mathcal{B}(t)$:
\begin{align}\label{Laplacetraf}
\X(\delta) \approx B_N(\delta) =\int_{0}^\infty  \!\! dt \e^{-t} t^{\,\beta_0} \mathcal{B}_N(t\delta)\,.
\end{align}
A straightforward approach to construct $\mathcal{B}_N(t)$ is the Pad{\'e}-Borel method, which uses Pad{\'e} approximants matched to the Borel series [Eq.~\eqref{Borelseries}].
[We note that while the conjectured large-order behavior ${a=-1/\pi}$ (and ${\beta=0}$)~\cite{Mari_o_2019}
would imply a non-Borel summable series for $\delta<0$, 
we find that the Pad{\'e}-Borel $\mathcal{B}_N(t\delta)$ approximants for $N\leqslant 4$ 
have no poles on the positive $t$ axis for $\delta<0$.]

In Fig.~\ref{fig:borel}, we show the results for $\X(\delta)$ from the \text{second-,} third-, and fourth-order 
Pad{\'e}-Borel approximants obtained using the standard choice ${\beta_0=0}$ for the Borel transform.
Also shown are the corresponding 
perturbative results as well as the results from QMC computations from the left 
panel of Fig.~\ref{fig:plot}.
One sees that, compared to the perturbative results, the Borel approximants for $\X(\delta)$ have a much better convergence behavior for $\delta>0.5$. 
Moreover, for ${|\delta| \lesssim 1}$ the fourth-order Borel results
are very close to the QMC data.

Overall, the results depicted in Figs.~\ref{fig:plot} and \ref{fig:borel} show that Pad{\'e} and Borel resummation methods 
allow to improve the convergence behavior of the $C_0$ part of the Fermi-momentum expansion.
To investigate this further
would require future computations of higher-order series coefficients beyond fourth order.
As discussed below, this however faces serious challenges.

\subsection{Beyond fourth order}\label{chap63}

The first complication regarding the calculation of coefficients beyond fourth order
is the rapid increase of the number of Hugenholtz diagrams with $N$.
Graph theory methods allow one to automatically
generate diagrams~\cite{stevenson,Arthuis:2018yoo,Arthuis:2020tjz}, 
from which one finds that the number of diagrams without single-vertex loops 
increases as $(1,1,3,39,840,27300,\ldots)$ for $N=(1,2,3,4,5,6,\ldots)$, where
the number of the relevant normal diagrams 
increases as $(1,1,3,33,668,21572,\ldots)$.

For a given set of higher-order diagrams, the evaluation of those 
without UV divergences and those that have only simple ladder-type divergences (which are renormalized by two-body counterterms)
would be relatively straightforward.
That is, for a given diagram the only complication compared to a fourth-order diagram of similar type would be additional three-momentum integrals.

The main challenge concerning higher-order calculations lies (as in the fourth-order case) with UV divergences 
that are not renormalized by two-body counterterms.
For instance, at fifth order one encounters several three-body scattering diagrams of the form 
of diagrams of Fig.~\ref{fig_3to3} but with two additional intermediate states.
These diagrams have logarithmic subdivergences that cancel if the diagrams are summed.
The remaining linear UV divergence cancels for $g=2$ and is otherwise renormalized by a momentum-independent three-body counterterm.
For the next diagonal Pad{\'e} approximant (Pad{\'e}$[3,3]$) one would have to go to sixth order, 
where a much larger number of diagrams with complementary subdivergences and
also the first momentum-dependent logarithmic divergence $\sim Q^2\ln(\Lambda/Q)$ appears in three-body scattering (see Sec.~\ref{chap23}).

\section{Summary}\label{chap7}

In this paper we have discussed high-order perturbative EFT calculations for fermions at very low energy scales. 
In particular the issue of renormalization has been investigated in detail.
We have then elaborated and expanded on our recent calculation~\cite{kfas1} of the fourth-order 
term in the Fermi-momentum or $\kF a_s$ expansion for the ground-state energy of the general dilute Fermi gas.
The result for the complete (i.e., including both analytic and logarithmic terms) fourth-order coefficient
has been given for two different regularization and renormalization schemes: 
cutoff regularization (with divergence subtraction) and dimensional regularization (with minimal subtraction).

The central results for the Fermi-momentum expansion are summarized in Sec.~\ref{chap5},
where in Table~\ref{ourtable}
the various contributions to the regular (i.e., nonlogarithmic) $(\kF a_s)^4$ part of the fourth-order term are listed.
In Sec.~\ref{chap6} we have then 
investigated the convergence behavior of the expansion for the case of spin one-half fermions.
Using Bayesian methods and comparing against results from nonperturbative QMC computations,
we found that the expansion is well-converged at fourth order for ${| \kF a_s | \lesssim 0.5}$, 
and exhibits divergent behavior for ${| \kF a_s | \gtrsim 1}$, see Fig.~\ref{fig:plot}.
(To be precise, the $\kF a_s$ expansion is a divergent asymptotic series;
by ``divergent behavior'' we mean that the accuracy of the result at low truncation orders is deficient.)

Furthermore, we have shown that 
Pad{\'e}-Borel resummations (of the $a_s$-only part of the expansion)
improve the convergence and give well-converged results at fourth order in the region ${| \kF a_s | \lesssim 1}$,
see Fig.~\ref{fig:borel}. Accurate results throughout the entire BCS regime with negative $\kF a_s$
(and into the BEC region) can however be obtained via resummations that incorporate constraints on the behavior for $\kF a_s\rightarrow -\infty$ from QMC computations~\cite{Wellenhofer:2020ylh,ODME2}.
Given the technical challenges that arise beyond fourth order, it is 
unlikely that the $\kF a_s$ expansion will be evaluated to even higher precision in the near future.

Our results for the Fermi-momentum expansion at fourth order provide important constraints for ultracold atoms and dilute neutron matter.
Specifically, our results serve as useful benchmarks for future QMC simulations of dilute Fermi systems
and may be used to construct improved models of neutron -star crusts.
Future work may be targeted at high-order calculations of the dilute Fermi gas expansion at finite temperature.

\begin{acknowledgments}
We thank 
T.~Duguet, H.-W.~Hammer, J.A.~Melendez, and S.~Wesolowski for useful
discussions, and P. Arthuis for sending us his list of fifth- and sixth-order diagrams. 
We are also grateful to S.~Gandolfi and S.~Pilati for sending us
their QMC results. This work is supported in part by  the  Deutsche
Forschungsgemeinschaft (DFG, German Research Foundation) -- \mbox{Project-ID}
279384907 -- SFB 1245, the US Department of Energy, the Office of Science, the
Office  of  Nuclear  Physics,  and  SciDAC under awards \mbox{DE-SC00046548} and
\mbox{DE-AC02-05CH11231}. C.D. acknowledges  support by the  Alexander von
Humboldt Foundation through a Feodor-Lynen Fellowship. This material is based upon work supported by the U.S. Department of Energy, Office of Science, Office of Nuclear Physics, under the FRIB Theory Alliance award DE-SC0013617.  Computational resources
have been provided by the Lichtenberg high performance computer of the TU
Darmstadt.
\end{acknowledgments}

\bibliography{refs}

\end{document}